\documentclass[tomography,article,preprint]{Definitions/mdpi}
\firstpage{1} 
\pubvolume{12}
\issuenum{1}
\articlenumber{49}
\pubyear{2026}
\copyrightyear{2026}
\externaleditor{Emilio Quaia} 
\datereceived{23 December 2025} 
\daterevised{19 March 2026} 
\dateaccepted{24 March 2026} 
\datepublished{1 April 2026}  
\doinum{10.3390/tomography12040049}
\usepackage{makecell}        

\Title{Standardized Images and Evaluation Metrics for Tomography}

\Author{Anna Frixou 
 $^{1}$\orcidA{}, Theodoros Leontiou $^{2,3}$\orcidB{}, Efstathios Stiliaris $^{1,4}$\orcidC{} and Costas N. Papanicolas  $^{1,4,5,}$*\orcidD{}}

\AuthorNames{Anna Frixou, Theodoros Leontiou, Efstathios Stiliaris and Costas N. Papanicolas}

\address{%
$^{1}$ \quad The Cyprus Institute,  2121 
Nicosia, Cyprus;   a.frixou@cyi.ac.cy (A.F.); stiliaris@phys.uoa.gr (E.S.) \\
$^{2}$ \quad Department of Mechanical Engineering, Frederick University,  1036 Nicosia, Cyprus;  t.leontiou@frederick.ac.cy \\
$^{3}$ \quad Frederick Research Center,  1036 Nicosia, Cyprus
\\
$^{4}$ \quad Department of Physics, National and Kapodistrian University of Athens, 15771  Athens, Greece
\\
$^{5}$ \quad  The Cyprus Academy of Sciences, Letters and Arts,  1522 Nicosia, Cyprus}

\corres{Correspondence:  c.papanicolas@cyi.ac.cy}

\simplesumm{Modern tomography methods produce high-quality images, making it difficult to evaluate and compare their performance using traditional image quality metrics. Many commonly used metrics tend to saturate and provide limited sensitivity to residual differences between reconstruction methods.
In this work, we introduce a structured evaluation framework based on standardized reference images and diagnostic tools. Central to the approach is the use of an ``Ideal Image'' as a physically meaningful benchmark and the Structure and Contrast Index (SCI) to quantify structured residual information. Using simulation studies in emission tomography (SPECT), we demonstrate how this framework supports algorithm development, benchmarking, and quality assurance.} 

\abstract{\textbf{Background/Objectives:} Modern tomographic reconstruction methods---including physics-informed and AI-based approaches---can produce very high fidelity images. In this regime, widely used global image quality metrics often approach saturation, making it harder to distinguish residual differences between methods and identify remaining performance gaps. This study introduces a physically grounded and standardized evaluation framework designed to retain sensitivity beyond conventional global metrics and support both comparison and systematic improvement in tomographic reconstruction methods.
 \textbf{Methods:} The proposed framework defines standardized reference images---``Source'', ``Detector'', ``Ideal'', and ``Realistic''---using Monte Carlo simulations, with the Ideal Image serving as a physically grounded benchmark. Reconstruction performance is evaluated using pixel-wise difference and $\chi^2$ maps, Region-of-Interest analysis, intensity (gray-value) histogram comparisons, and the Structure and Contrast Index (SCI), computed on difference maps. Demonstrations use simulated SPECT data reconstructed with ART, MLEM, and RISE-1.
 \textbf{Results:} Across case studies, SCI and $\chi^2$-based diagnostics reveal structured residuals and localized deficiencies not evident from global similarity metrics such as SSIM or NMSE. Comparative analyses show that methods with similar global scores can exhibit distinct residual structures and region-specific performance variations, while improved agreement in the sinogram domain does not necessarily translate into improved image fidelity. Histogram-based diagnostics provide complementary information on intensity redistribution not captured by pixel-domain summaries.
 \textbf{Conclusions:} The framework provides a reproducible, physically meaningful, and sensitive approach for evaluating tomographic reconstruction performance in the high-fidelity regime. By combining standardized reference images with multi-domain and multi-metric analysis, it enables robust benchmarking and supports physically consistent interpretation of reconstruction quality.
}


\makeatletter
\let\old@linenumbers\linenumbers
\def\linenumbers{} 
\makeatother

\begin{document}
\section{Introduction} 

The quality of tomographic image reconstruction is of fundamental importance in a wide range of fields---from clinical diagnostics in nuclear medicine to structural imaging in industrial and materials science. In recent years, the field has experienced dramatic advances, driven by both improved detector instrumentation and the increasing availability of powerful computational resources. These developments have enabled the use of sophisticated reconstruction methods, including algorithmic, AI-based, and hybrid physics-informed approaches, that operate with high precision and accuracy. The shift toward neural-network-driven solutions and the inherent need to detect ``hallucinations'' or structural artifacts in high-performance regimes~\cite{9424044, KUMARI2024107912, Adler_2017, Chen2023_unpaired_I2I, Jiang2025_PMB, MacaronePalmieri2025PhysScr, XuNoo2022PMB, Zeng2022PMBCTP}. Modern hybrid approaches enable the incorporation of prior physical knowledge into the reconstruction process, leading to physically possible solutions instead of pure mathematical ones~\cite{a17020071, Han2025PhysicsInformedDiffusionCT, Xie2021AnatomicallyAidedPET}. As a result, reconstruction techniques are now approaching the limits of what is physically achievable, although the extent of this gap has yet to be clearly quantified. 

However, this progress also exposes a critical limitation in how reconstruction performance is currently evaluated. Tomographic image reconstruction is a mathematically ill-posed inverse problem:   
for any given set of projection data (e.g., a sinogram), multiple images may exist that are consistent with it. Reconstruction methods differ in how they incorporate assumptions, regularization, or prior knowledge to resolve this ambiguity. Yet the tools typically used to assess reconstruction quality---simple digital or hardware phantoms and global metrics such as the Structural Similarity Index Measure (SSIM), Normalized Mean Squared Error (NMSE), correlation coefficient (CC), or Peak Signal-to-Noise Ratio (PSNR)---were not designed for this high-performance regime. These traditional metrics tend to saturate in high-quality reconstructions, making it difficult to capture subtle but diagnostically relevant differences, thereby offering limited guidance for further methodological development. Recent efforts tried to establish digital phantoms~\cite{Ghaly_2014, harries2020realistic} and evaluation criteria across different modalities like Computed Tomography (CT) and Positron Emission Tomography (PET), but there is still work to be done.

Furthermore, evaluation is often based on a single reference image---typically the ``phantom''---which is assumed to represent ground truth. While this approach is widespread, it is conceptually problematic. The phantom defines the object to be imaged, but its detailed structure cannot be reconstructed without prior knowledge, compounding the ill-posedness of the problem. Physical processes such as scattering, attenuation, noise, and imperfect detection introduce distortions that make exact recovery of the true distribution---represented by what we term the ``Source Image''---fundamentally impossible. As a result, comparisons made directly against the ``Source Image'' may lead to misleading conclusions about reconstruction method performance. Instead, we argue that multiple reference images are needed---including an ``Ideal'' image that represents the best possible reconstruction achievable under perfect acquisition conditions, and it is the one that should be used for evaluation purposes of reconstruction algorithms.

These 
 challenges raise a set of fundamental questions:
\begin{itemize}
    \item What set of reference images should be used to evaluate a reconstruction method in a meaningful and physically realistic way?
    \item How can we distinguish between methods that perform similarly under traditional metrics?
    \item How can we detect local deficiencies or residual information when global scores plateau?
    \item And is it possible to quantify the gap that exists between a given reconstruction and what is physically achievable?
\end{itemize}

Together
, appropriately chosen phantoms and refined quantitative metrics form a robust, structured, and reproducible framework for assessing the performance of reconstruction methods in a physically meaningful way. We demonstrate the utility of this framework through detailed case studies involving the well-established Maximum Likelihood Expectation Maximization (MLEM) and Algebraic Reconstruction Technique (ART) algorithms and the novel Reconstructed Image from Simulations Ensemble-1 (RISE-1) method, applied to the Shepp--Logan software phantom. While our focus is on Single-Photon Emission Computed Tomography (SPECT), the structure of the framework extends naturally to PET with only minor adjustments to the attenuation and coincidence models. Its application to other modalities---such as CT, Magnetic Resonance Imaging (MRI), or thermal imaging---is possible in principle but requires modality-specific definitions of the standardized images and careful adaptation of the underlying forward and noise models. In what follows, we describe the rationale, construction, and application of each element in this evaluation toolkit.

\section{Materials and Methods}
The overarching goal of tomographic image reconstruction is to obtain a 3-D image depicting aspects of the imaged object as accurately as possible from the obtained experimental data (most often organized in the form of a sinogram). As the tomographic image reconstruction is mathematically an ``ill-posed problem''~\cite{a17020071, 9424044}, the challenge is to select among the many possible solutions that the sinogram can support, often injecting prior knowledge into the analysis~\cite{9424044}.

The use of both software (digital) and hardware phantoms plays a vital role in the development and refinement of image reconstruction methods~\cite{712135,Ghaly_2014,harries2020realistic,pullens2010ground}. The use of phantoms offers a most valuable advantage: the reconstructed image can be directly compared to the known structure of the object being imaged. This in turn necessitates the development and use of metrics that can quantify this comparison. In our efforts to develop new methodologies and enhance existing ones, we found that while established metrics were effective for certain studies, the introduction of new standardized images and phantoms and of new evaluation criteria and methods was particularly valuable. The newly introduced images and metrics reported in this work have broad applicability, which is showcased in \mbox{two examples:}
\begin{enumerate}[label=(\alph*)] 
\item In evaluating improvement and convergence in the well-known case of the MLEM reconstruction algorithm~\cite{Hwang_2006}; 
\item In evaluating the performance of a novel reconstruction algorithm and benchmarking it versus the well-known ART and MLEM. 
\end{enumerate}

\subsection{Phantoms}

In emission tomography, a phantom---whether realized in software or as a physical object---refers to a well-defined arrangement of radioactive emitters (e.g., technetium-99m, $^{99m}Tc$) that serve as the source of radiation detected by the imaging system (such as $\gamma$-rays in SPECT).  This arrangement is characterized by a precisely specified spatial distribution, in two or three dimensions, together with the emission intensity (specific activity) at each location. Invariably, phantoms also include detailed descriptions of surrounding materials---both within and outside the emitting regions---including density, absorptivity, and other properties relevant to the physics of $\gamma$-ray propagation. The shape and physical properties, even of non-emitting $\gamma$-ray structures (e.g., bone), influence the attenuation and scattering of the emitted $\gamma$-rays before they reach the detector.

In this work, we use the term phantom exclusively for this physical or digital configuration to clearly distinguish it from its ``Source Image''---the graphical representation of the radiation-emitting portion of the phantom. Other image-based representations derived from this primary configuration are introduced in Section \ref{sec:need_im}.

\begin{figure}[H]
     
    \includegraphics[width=0.95\linewidth]{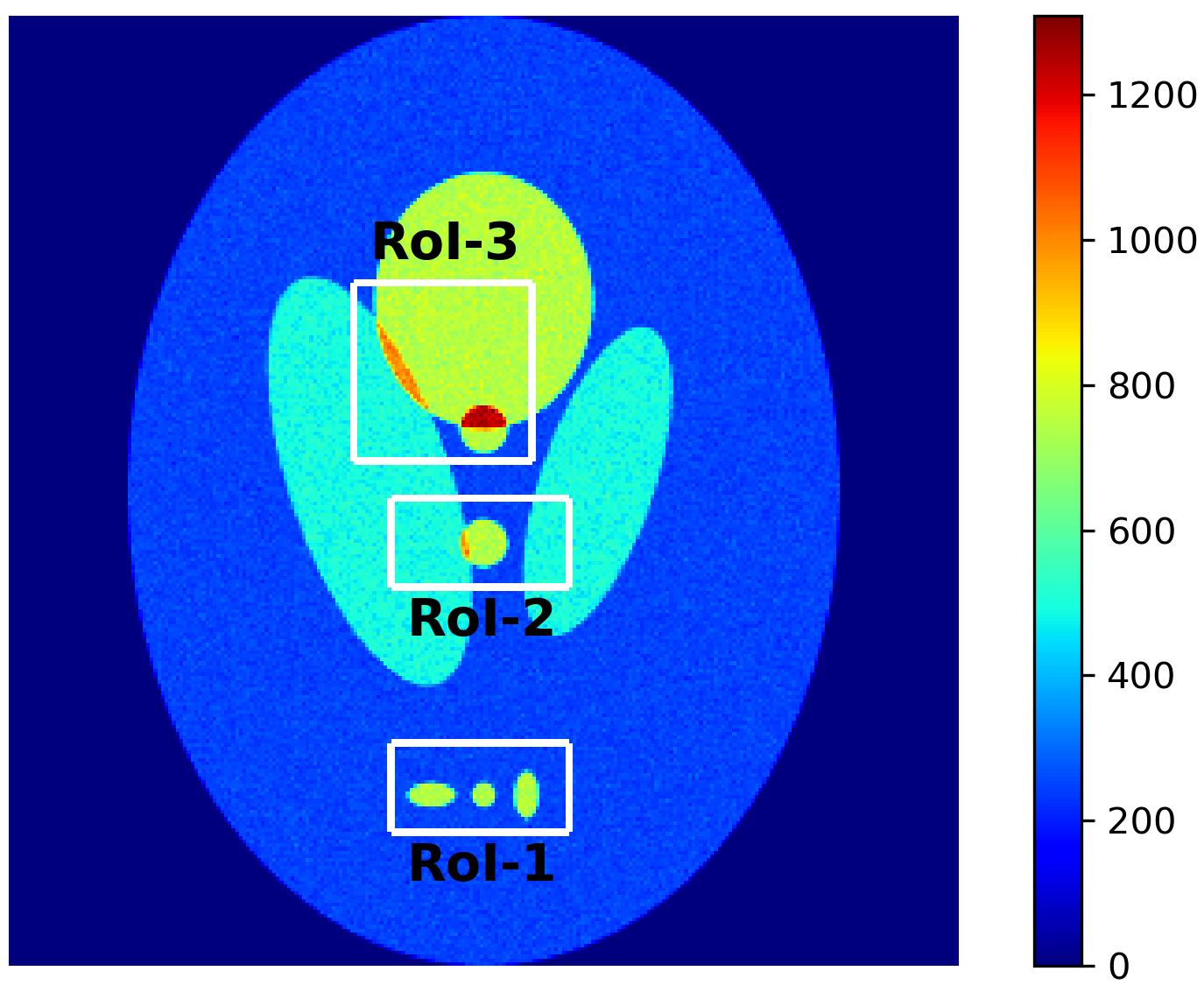}
    \caption{Graphical representation of the specific activity of the two-dimensional ``Source Images'' of the  Shepp--Logan phantom. Selected Regions of Interest (RoIs) are shown and will be discussed at Section~\ref{sec:roi_def}.} 
    \label{fig:SL_roi}
\end{figure}

For our investigation, we employed several phantoms, both standardized and custom-designed. In the results presented here, we focus on a widely used software phantom: the Shepp–Logan~\cite{shepp1974fourier, 4616690} phantom which is shown graphically in Figure~\ref{fig:SL_roi}. Variations of it are used throughout the imaging community. Extending the analysis to other phantom designs or to three-dimensional cases is conceptually straightforward. The Shepp–Logan phantom extends evaluation to heterogeneous, anatomically inspired structures. 
In the modified Shepp--Logan phantom version used in this work, the cranial bone has been removed, while the major large ellipsoid representing the brain grey and white serves also as a background source. Geometrical details and source activities of the two variants  of the modified Shepp–Logan phantom used in this work are shown in Table~\ref{tab:shepp_logan}. SRC0 is the background water source for each of the two phantoms. The first variant has a moderate background source (first line of Table~\ref{tab:shepp_logan}) while the second has a four times higher value (second line of Table~\ref{tab:shepp_logan}); 
SRC3–SRC8 have the same and two times the specific activity of the two bigger ellipsoids SRC1 and SRC2. Table~\ref{tab:shepp_logan} summarizes the geometric and activity parameters of the modified Shepp–Logan phantom.

The specific activity distributions of the Shepp–Logan phantom is shown graphically in Figure~\ref{fig:SL_roi}.

\begin{table}[H]

\caption{Description 
 of the radiation-emitting areas of the Shepp–Logan phantom consisting of eight ellipsoids (SRC1–SRC8) of various sizes and activities. The second phantom configuration, used for the MLEM-RISE-1 comparison, features a background activity that is four times higher (see line 2 of the table), representing a more demanding reconstruction scenario.\label{tab:shepp_logan}}

\begin{adjustwidth}{-\extralength}{0cm}
 
\begin{tabularx}{\fulllength}{LCCCCCCCC}
\toprule
\textbf{Source} 
&  \textbf{Position [mm]}  
& \textbf{Minor X Axis [mm]}  
& \textbf{Major Y  Axis [mm]}  
& \textbf{Angle \boldmath{$\phi$}  [deg]}  
&  \textbf{Surface/\boldmath{$\pi$} [mm\boldmath{$^2$}]}  
&  \textbf{Surface [a.u.]}  
& \textbf{Activity [kBq]} 
& \textbf{Specific Activity [kBq/a.u.]}  \\
\midrule
SRC0 & (0, 0)      & 75   & 100  & 0$^\circ$    & 7500.00 & 1200 & 6000 & 5 \\
  &              &        &        &                  &             &             & 24,000 & 20 \\
SRC1 & (+24, +2)   & 12   & 34   & $-$18$^\circ$  & 408.00  & 65   & 325  & 5 \\
 SRC2 & ($-$24, +2)   & 17   & 45   & +18$^\circ$  & 765.00  & 122  & 610  & 5 \\
 SRC3 & (0, +40)    & 23   & 27   & 0$^\circ$    & 621.00  & 100  & 1000 & 10\\
 SRC4 & (0, +13)    & 5    & 5    & 0$^\circ$    & 25.00   & 4    & 40   & 10 \\
 SRC5 & (0, $-$11)    & 5    & 5    & 0$^\circ$    & 25.00   & 4    & 40   & 10 \\
 SRC6 & ($-$11, $-$64)  & 5    & 2.5  & 0$^\circ$    & 12.50   & 2    & 20   & 10 \\
SRC7 & (0, $-$64)    & 2.5  & 2.5  & 0$^\circ$    & 6.25    & 1    & 10   & 10 \\
 SRC8 & (+9, $-$64)   & 2.5  & 5    & 0$^\circ$    & 12.50   & 2    & 20   & 10 \\
\noalign{\hrule height 1pt}
\end{tabularx}
\end{adjustwidth}
\end{table}
    
\subsection{{Evaluation of Reconstruction Algorithms in Medical Imaging}}
Reconstruction algorithms across a range of imaging modalities are routinely assessed both in absolute terms and through comparative analysis, with a focus on their fidelity in reproducing the underlying imaged object. These evaluations are typically performed using a combination of physical and digital (software) phantoms, which serve as “ground-truth” references. The reconstructed images are quantitatively compared to the known structure of these phantoms using a suite of established and well-characterized metrics. These metrics are designed to quantify various aspects of image quality, such as spatial resolution, contrast recovery, and noise characteristics. In Section~\ref{sec:results2}, we will explore in detail the extent to which these metrics effectively capture the reconstruction performance of a given methodology and how reliably they discriminate between alternative \mbox{reconstruction methods.}

Despite the apparent maturity of this evaluation framework, several important questions remain open. For instance, is it theoretically possible for any reconstruction methodology to perfectly reproduce a source phantom given ideal conditions? If not, what fundamental limitations---whether physical, computational, or methodological---define the closest achievable approximation? Furthermore, when two reconstruction methodologies yield comparably high metric scores, it remains unclear whether this indicates genuine equivalence in performance, a saturation point dictated by physical constraints, or merely a shortcoming in the sensitivity of the employed evaluation metrics. It is plausible that both limitations of the metrics and physical asymptotes in performance play a role. In the subsequent sections, we aim to investigate these questions and offer at least partial answers supported by theoretical insights and empirical evidence.

Central to this inquiry is the availability of highly advanced simulation frameworks that enable precise modeling of the underlying physical processes in imaging. In particular, the Monte Carlo-based GEANT4/GATE software suite~\cite{jan2004gate,  sarrut2022opengate} has emerged as an indispensable tool in nuclear medicine imaging modalities. Its capability to simulate complex interactions of particles or photons with matter at high accuracy, combined with the increasing affordability of intensive computational resources, provides a powerful foundation for probing the ultimate limits of image reconstruction accuracy in SPECT and PET modalities. These tools allow for the generation of controlled, high-precision datasets, which are critical for disentangling methodological limitations from intrinsic physical constraints.

For the Shepp–Logan phantom, the SPECT image results of the simulation presented in this paper were produced using the Monte Carlo GEANT4/GATE~\cite{jan2004gate,  sarrut2022opengate} software package together with a typical $\gamma$-camera dual-head system routinely used in a clinical environment and radio-tracer $^{123}\mathrm{I}$ with $\gamma$-rays at159 keV. The phantoms are constructed using the simple analytical functions allowed by GATE, with the characteristics shown in Table~\ref{tab:shepp_logan}. Since our study focuses on the reconstruction quality of the central-slice image only, a natural thickness of $\pm$5 mm is given to the phantom geometry. A total of 128 planar images are acquired with this simulated $\gamma$-camera system, corresponding to an angular step of $\Delta \theta$ = 2.8125$^\circ$, covering the full 360$^\circ$  range. The obtained phantom matrices and the generated sinograms are normally composed to a 256 × 256 pixel image size, unless otherwise specified.

\subsection{{The Need for Standardized Images}}\label{sec:need_im}
While the foundational principles of tomography were established~\cite{4307775} and significant early advances were achieved~\cite{seynaeve1995historiek} in an era predating modern computing, the ability to incorporate detailed, physics-based modeling of all relevant processes influencing image formation was severely limited. Early methodologies often relied on drastic simplifying assumptions, as the computational tools necessary for realistic simulations did not yet exist.  Today, however, the field benefits from highly accurate and comprehensive simulations, capable of modeling even subtle effects that impact the quality of the reconstructed image. In the case of emission tomography, these include inhomogeneities in absorption due to variations in the material between the source and detector, scattering phenomena, and differences in imaging geometry or detector response.

As a result, the reconstructed image---and its corresponding sinogram---can vary depending on the specific assumptions and physical models used in the simulation. Rather than treating this variability as a limitation, it offers an opportunity: these simulation-derived variants can serve as new reference standards against which to compare the outputs of reconstruction methodologies under evaluation. Such comparisons provide a more nuanced and realistic benchmark for methodology performance, grounded in the complex physical realities of image formation. The set of standardized images required for this purpose is presented below only for emission tomography. By analogy, the same conceptual images can be defined for other modalities (e.g., MRI and CT); however, a different physical forward model must be applied, incorporating the specific constraints of \mbox{each system.}

\begin{figure}[H]
     
    \includegraphics[width=1.0\linewidth]{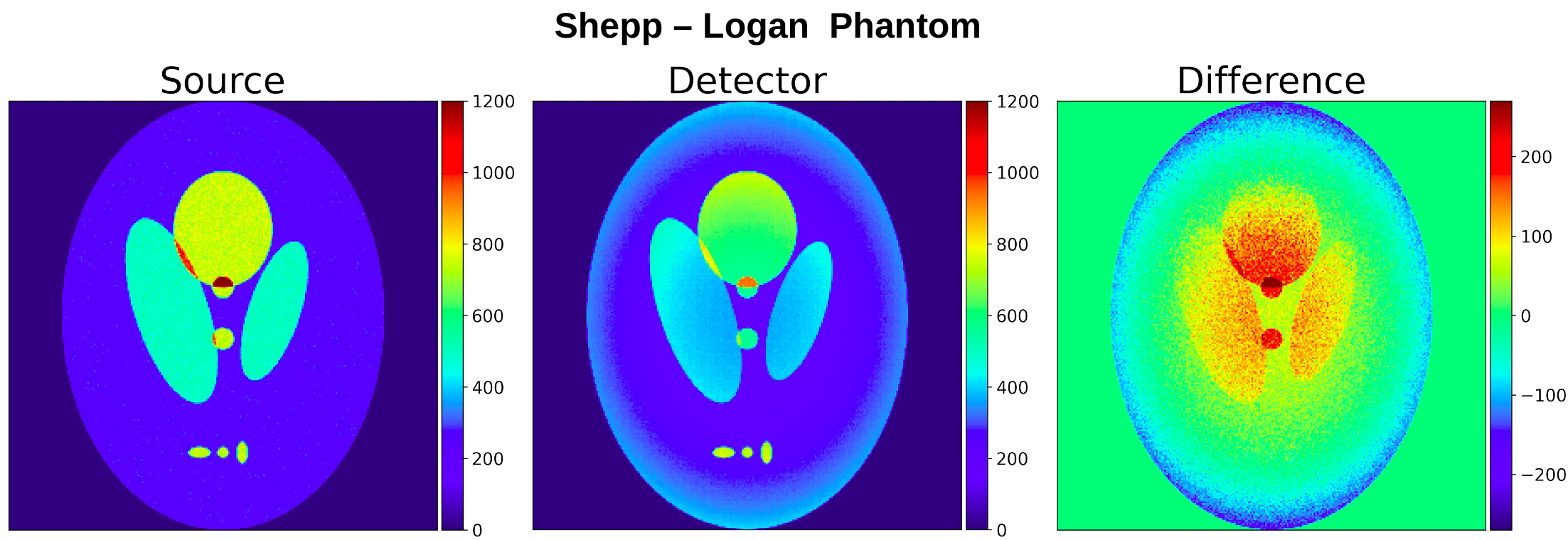}
    \caption{
    	``Source'' and ``Detector'' images from simulations of the modified  Shepp–Logan phantom. The ``Source Image'' includes all emitted photons; the ``Detector'' image includes only detected ones. The rightmost panels show their difference; photon absorption and scattering within the phantom is the dominant effect.}
    \label{fig:detector}
\end{figure}

A simple image that is produced by simulating in detail the processes that occur from emission to detection is what we term the ``\textbf{Detector Image}
'', which is constructed by using the nominal position of the emitted photons that were detected.  Figure~\ref{fig:detector} shows the images of the source phantom and of the ``Detector'' image for the test phantom used in our study and the difference among them. This image is useful for forward-model validation and field-application studies.

A widely used reconstructed image obtained using detailed simulation packages is the image that is expected to emerge from a realistic detection system---e.g., a commercial scanner. In such simulations the detailed specification of a digital or hardware source phantom and every detail of the detection system is meticulously described. This image, which we term the ``\textbf{Realistic Image}'', is particularly useful for field applications---for example, in a clinic or a laboratory that relies on a specific detection setup and equipment to derive its results. The ``Realistic Image'' is the appropriate tool in such environments to explore issues such as optimization of radiopharmaceutical dose (in PET or SPECT), collimator geometry (in SPECT), number of planar images to be taken, and similar parameters. This image is directly related to the hardware and experimental conditions, which make it unsuitable for assessing the capabilities and performance of a reconstruction methodology, as discussed above. To facilitate this need we propose the introduction of the ``\textbf{Ideal Image}''.
In practical imaging scenarios, no matter how advanced the reconstruction methodology is, the resulting image will never reach a statistically perfect representation of the source phantom. This is due to several limiting factors:
\begin{enumerate}[label=(\alph*)]
    \item The interaction of the emitted photons within the body containing the volume to be imaged (e.g., a patient’s body), including scattering, annihilation, or kinematic blurring (in PET), some of which are often grouped under the encompassing term \textit{attenuation
}.
    \item The interaction of the photons with intervening material between the boundaries of the imaged object and the detector array (e.g., air, collimator, and encasing materials in SPECT).
    \item Limitations of detection such as finite angular acceptance or energy resolution.
    \item Limitations in data acquisition, including a finite number of projections and limited counting statistics.
    \item Electronic noise and uncertainties in the instrumentation chain leading to event recording.
    \item Uncertainties introduced in analyzing finite-statistics digitized data (e.g., binning errors).
\end{enumerate}

With 
 the exception of case (a), these physical limitations can be mitigated by acquiring a sufficiently large number of projections and allowing for extended counting times to ensure robust statistics. The image resulting from such a simulation---conducted under these near-ideal conditions---is what we define as the ``Ideal Image''. It represents the highest level of accuracy attainable and differs from the ``Source Image'' solely due to the fundamental photon interactions (attenuation and scattering) simulated within the GATE framework. This image serves as an absolute benchmark against which practical reconstruction methods can be evaluated. In this sense, the ``Ideal Image'' is universal: it is independent of specific detection geometry, equipment, or algorithm and is determined solely by the fundamental physics and statistical constraints of the imaging system.

\begin{figure}[H]
     
    \includegraphics[width=1.0\linewidth]{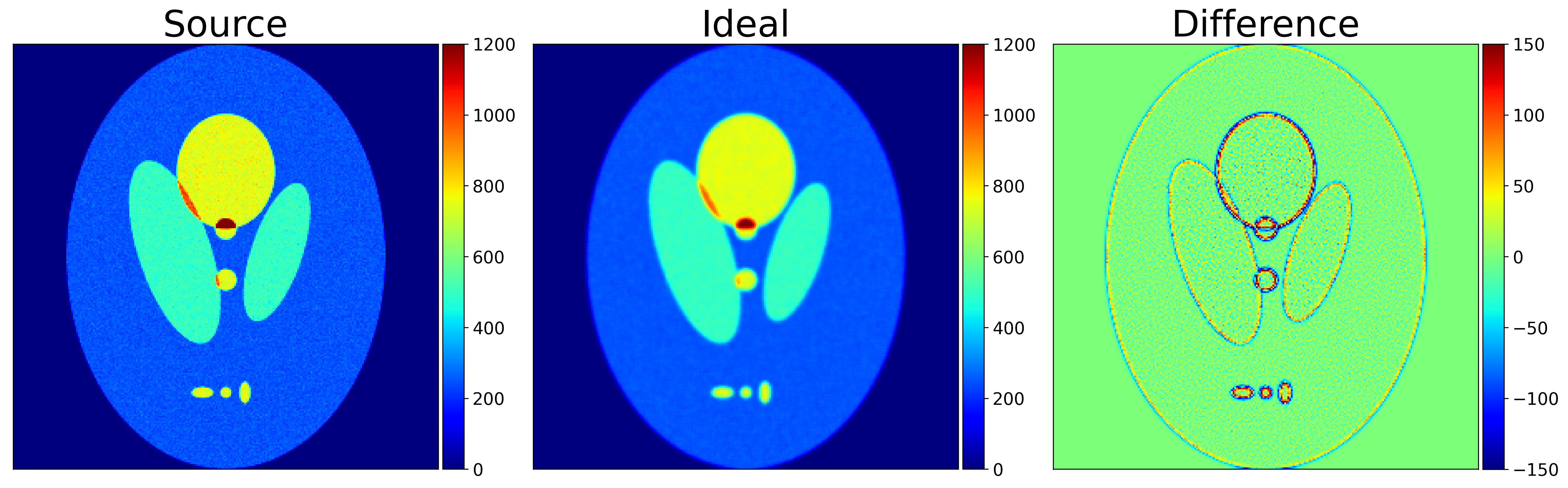}
    \caption{Simulated Shepp--Logan image of the “Source Image” and the “Ideal Image”, together with their difference. }
    \label{fig:dif_ideal}
\end{figure}

In order to address the issue of minimizing the effects deriving from the geometric and construction peculiarities of the collimator---a very important consideration in SPECT due to finite angular acceptance---the concept of the ``\textbf{Ideal Collimator}'' was introduced. In imaging systems like SPECT, collimators are made of dense, absorbing materials that allow for only rays traveling in specific directions to reach the detector. However, collimators are not infinitely narrow, which introduces blurring: some rays scatter on or penetrate through the collimator and reach neighboring detector channels. In simulations, a software collimator ignores the finite thickness of a hardware collimator and allows for each channel to collect rays from a single direction. An ``Ideal Collimator'' is therefore effectively a software filter that accumulates only rays emitted perpendicularly to the detector surface. Differences between the image of the phantom and the ``Ideal Image'' thus reflect inherent deterioration of the image due to (a) above, which no reconstruction methodology can overcome without the use of prior-knowledge input. Such a comparison is shown in Figure \ref{fig:dif_ideal}. Differences between the image derived through a reconstruction methodology and its corresponding ``Ideal Image'' are attributable to deficiencies in the reconstruction methodology. Quantifying these differences or deficiencies will be the topic of the following section. Table~\ref{tbl:4im} summarizes the four types of images we introduced, along with the method of generation and principal utility.

\vspace{-6pt}
\begin{table}[H]
\caption{The 
 four images and sinograms defined and used in this work. \label{tbl:4im}}
 
\begin{tabularx}{\textwidth}{lLL}
\toprule
\textbf{Image/Sinogram} & \textbf{Generation Method} & \textbf{Comments} \\
\midrule
``Source'' 
 & Direct emissivity map of the phantom. No attenuation, scatter, or detector effects; no forward model needed. & Provides the reference “reality” and a benchmark for testing reconstruction methodologies. \\
\midrule
``Detector'' & Forward model accounts for attenuation, scatter, geometric acceptance, and detection efficiency. Provides an excellent representation incorporating all detection physics. & Represents the starting point for any reconstruction; reconstructed images should improve upon the ``Detector'' image. \\
\midrule
``Ideal'' & Forward model under asymptotic conditions: infinite statistics, ``Ideal Collimator'', and no detector blur. Represents the physical upper bound on recoverable information. & Provides an absolute benchmark for testing reconstruction methodologies.\\
\midrule
``Realistic'' & Forward model with full system effects: finite resolution, collimator blur, detector response, and realistic counts. Represents actual scanner behavior. & Useful for field applications (e.g., clinics or laboratories) that rely on specific detection equipment to derive results. \\
\bottomrule
\end{tabularx}
\noindent \footnotesize{The method of their generation and their principal utility is listed in the table.  They are discussed in a more thorough manner in the text.}

\end{table}

This taxonomy is not merely conceptual: it follows directly from the structure of the physical forward model used in SPECT and other tomographic modalities. The ``Source Image'' represents the emissivity term; the ``Detector Image'' represents, with high fidelity, the emission and detection physics underlying the process; the ``Ideal Image'' corresponds to the forward model stripped of instrumental imperfections and noise; and the ``Realistic Image'' incorporates all real system effects. These four images therefore isolate distinct physical contributions---source distribution, propagation physics, detection characteristics, and acquisition statistics---and are essential for disentangling algorithmic limitations from physical constraints. In this sense, the taxonomy is required for any rigorous, physics-grounded evaluation of reconstruction fidelity.

\subsection{{Quantitative and Diagnostic Evaluation Tools}}
As outlined in the preceding discussion, the development and validation of novel image reconstruction methodologies necessitate the use of well-characterized, standardized benchmarks, along with robust and meaningful quantitative metrics. Such benchmarks are essential to ensure objective, reproducible, and interpretable evaluations of performance.
The standardized images defined in Section \ref{sec:need_im}---``Source'', ``Detector'', ``Ideal'', and ``Realistic''---provide the physically meaningful anchors needed to evaluate fidelity across image and sinogram domains. These images establish a structured framework against which reconstructed outputs can be compared.
In the present section, we extend the evaluation methodology by
\begin{enumerate}[label=(\roman*)]
    \item Recalling the global quantitative metrics that remain essential for baseline comparisons;
    \item Introducing local and histogram-based diagnostics that retain sensitivity when global measures saturate; 
    \item Formalizing Region-of-Interest (RoI) analysis as a focused evaluation tool.
\end{enumerate}

The 
 section concludes with an integrated interpretation framework that unifies all these approaches.

In Table~\ref{tbl:metrics}, metrics and evaluation criteria used in our investigations are presented. All methods are applicable to both software and hardware phantoms. For the well-established ones, references are provided, while those introduced in this work  are marked \mbox{as “new”.}

\begin{table}[H]
 
\caption{Applicability 
 of evaluation metrics and diagnostic tools across different imaging contexts. \label{tbl:metrics}}
  
\begin{tabularx}{\textwidth}{m{7.8cm}<{\raggedright}m{3cm}<{\raggedright} m{2cm}<{\raggedright}}
\toprule
\textbf{Metric} & \textbf{Field Data} & \textbf{References} \\
\midrule
Correlation Coefficient (CC) & Only on sinograms &~\cite{hill2001registration, Schober2018} \\
\midrule
Normalized Mean Square Error (NMSE) & Only on sinograms &~\cite{wang2009mse} \\
\midrule
Peak Signal-to-Noise Ratio (PSNR) & Only on sinograms &~\cite{hore2010psnr} \\
\midrule
Contrast-to-Noise Ratio (CNR) & Only on sinograms &~\cite{geissler2007cnr} \\
\midrule
Structural Similarity (SSIM) & Only on sinograms &~\cite{995823} \\
\midrule
Structure and Contrast Index (SCI) &  \checkmark & new \\
\midrule
$\chi^2$ maps of sinograms &  \checkmark & new * \\
\midrule
$\chi^2$ maps of images &  \texttimes & new *\\
\midrule
Difference map of sinograms and SCI & \checkmark & new * \\
\midrule
Difference map of images and SCI & \texttimes & new *\\

\bottomrule
\end{tabularx} 
\end{table}

\begin{table}[H]\ContinuedFloat
 
\caption{{\em Cont.} \label{tbl:metrics}}
  
\begin{tabularx}{\textwidth}{m{7.8cm}<{\raggedright}m{3cm}<{\raggedright} m{2cm}<{\raggedright}}
\toprule
\textbf{Metric} & \textbf{Field Data} & \textbf{References} \\
\midrule

Sinogram intensity (gray-value) histogram analysis and $\chi^2_{reduced}$ of intensity histograms \vspace{0.2cm} & \checkmark & new * \\
\midrule
Image intensity (gray-value) histogram analysis and $\chi^2_{reduced}$ of intensity histograms \vspace{0.2cm} & \texttimes & new * \\
\bottomrule
\end{tabularx}
\noindent \footnotesize{(\checkmark ) indicates conditional or partial applicability; (\texttimes)   indicates the metric cannot be meaningfully applied in that setting. For ``field'' (experimental or clinical) data, image-based comparisons often require projections to sinogram space.
    Histogram-based diagnostics quantify differences in gray-value distributions and do not represent spatial-frequency or Fourier-domain analysis.
  *  These tools are widely used qualitatively, but in this work a subjective, quantitative formalization (SCI for difference maps, $\chi^2_{reduced}$ for $\chi^2$-maps, and $\chi^2_{reduced}$ for intensity-distribution comparison) is introduced.}

\end{table}

\subsubsection{{Global Quantitative Metrics}}\label{sec:gl_metr}
Several well-established quantitative metrics are routinely employed to assess the quality of reconstructed sinograms and images and their fidelity to reference data. These metrics are applicable to both reconstructed images and sinograms and are instrumental in benchmarking methodological performance.

Global metrics often condense image or sinogram agreement into a single, reproducible number such as $\chi^2$. They are easy to report and compare but can mask localized or structured residuals once reconstructions are already of high quality. We therefore use them as a baseline and subsequently probe deeper with the diagnostics presented in Sections~\ref{sec:local_tools} and~\ref{sec:roi_def}. For the sake of completeness, accuracy, and reproducibility, we provide both the references (see Table~\ref{tbl:metrics}) and the exact formulas in Appendix~\ref{app:formula}.

\begin{itemize}
\item[] \hspace*{-7.5mm}Global Scalar Metrics 

\end{itemize}

Global scalar metrics remain widely used in tomographic image analysis because they condense the agreement between two images or sinograms into a single reproducible quantity. In principle, many such metrics may be employed to evaluate reconstruction fidelity. Beyond the classical $\chi^2$ statistic, we have examined a broader family of global comparison measures, including information-theoretic divergences such as the Jensen–Shannon (JS) divergence~\cite{Lin1974, Endres2003} and the Kullback–Leibler (KL) divergence~\cite{Kullback1951}; statistical–geometric measures such as the Hellinger distance~\cite{LeCam1986}; and optimal-transport distances such as the 1‑Wasserstein and 2‑Wasserstein metrics~\cite{Villani2009, peyre2019computational, Flamary2021}. These modern alternatives often retain sensitivity in high‑fidelity regimes, where conventional scalar metrics (e.g., SSIM, NMSE, CC, and PSNR) tend to saturate and lose discriminative power.

\begin{itemize}
\item[] \hspace*{-7.5mm}On the role of metrics as cost functions.
\end{itemize}

Many of these comparison metrics are not only used for evaluation purposes but also frequently serve as cost functions guiding the reconstruction process itself. Iterative and model‑based algorithms often optimize a specific discrepancy measure---whether a likelihood‑based statistic such as the $\chi^2$, an information divergence such as the Kullback–Leibler divergence, or a geometric discrepancy arising from optimal transport. The choice of cost function thus strongly influences convergence behavior, noise propagation, and the types of structures preferentially preserved or suppressed. The optimal choice depends on the physical acquisition model, statistical noise characteristics, and diagnostic priorities of the imaging modality. In this respect, although more sophisticated metrics can offer improved sensitivity or robustness, simpler measures such as the $\chi^2$ retain practical advantages in transparency, physical grounding, and broad applicability across imaging and sinogram domains.  

\begin{itemize}
\item[] \hspace*{-7.5mm}Rationale for the use of $\chi^2$ in this work.
\end{itemize}

For the purposes of this manuscript, we have deliberately adopted the $\chi^2$ statistic as the primary global scalar metric. This decision reflects considerations of interpretability, methodological consistency, and reproducibility across all reconstruction scenarios examined in this study, rather than an assertion of its superiority over alternative metrics. The $\chi^2$ metric provides a physically grounded, noise-aware measure of discrepancy and can be uniformly applied in image, sinogram, and intensity histogram analysis within our evaluation framework.

Formally
, given two arrays ($X$) and ($Y$) representing reconstructed and reference data (image or sinogram) and corresponding variance estimates ($\sigma_i^2$), the global $\chi^2$ metric is defined as $\chi^2 = \sum_i \frac{(X_i - Y_i)^2}{\sigma_i^2}$, where the sum extends over all valid pixels or bins. When explicit variance estimates are unavailable, reduced or unweighted $\chi^2$ variants may be used, though their interpretation becomes less strictly statistical.

These global metrics, collectively, provide a necessary quantitative foundation for evaluating reconstruction methodologies. However, once their values saturate---especially for SSIM and CC---their discriminative power diminishes. The next subsections therefore introduce local and histogram-based diagnostics that remain sensitive in this regime.

In all $\chi^2$ evaluations presented in this work, the variance entering the denominator is derived directly from the Monte Carlo simulation and is therefore dominated by counting statistics. The underlying signal is taken as noise-free. The $\chi^2$ used throughout is the reduced $\chi^2$, computed either globally or pixel-wise to form $\chi^2$ maps.

Pixel independence cannot be assumed in sinograms or reconstructed images, since both contain spatially correlated structures. For this reason, $\chi^2$ is used as a discrepancy measure and as a reconstruction cost function but not for estimating parameter uncertainties; those are obtained with the Jackknife method. The choice of $\chi^2$ reflects simplicity and transparency, not exclusivity: other discrepancy metrics were examined, and the qualitative conclusions reported here do not depend on the particular cost function. As noted elsewhere, the optimal choice of discrepancy metric is modality- and application-dependent.

\subsubsection{{Local and Intensity (Gray-Value) Histogram Diagnostic Tools}}\label{sec:local_tools}

\begin{itemize}
\item[] \hspace*{-7.5mm}Motivation
\end{itemize}

Global metrics (Section~\ref{sec:gl_metr}) provide indispensable, reproducible scalar summaries of agreement between reconstructed and reference data. However, in the high-fidelity regime reached by modern methodologies, these metrics often reach asymptotic behavior (e.g., global $\chi^2$ values) or saturate (e.g., $\text{SSIM} \approx 1$; $\text{NMSE} \approx 0$), concealing where residual discrepancies remain and whether those residuals are random or contain structured information that can still be recovered.

To overcome this limitation, we employ a set of spatial and histogram-based diagnostics designed to (i) expose localized deviations, (ii) evaluate their statistical significance, and (iii) determine whether the residuals are organization-free (noise-like) or structurally coherent (information-bearing). These diagnostics are applied both in image and sinogram domains, using the standardized images introduced in Section~\ref{sec:need_im}.

\begin{enumerate}
\item \textbf{$\chi^2$ \hspace{0.05cm} and Image Difference Maps}\\ 
Image difference maps and $\chi^2$ maps serve as valuable diagnostic tools, providing spatially resolved insight into how reconstructed data compare with a reference image or sinogram. Difference maps highlight pixel (or voxel)-wise discrepancies, allowing structural mismatches to be identified in images and projection-space inconsistencies to be detected in sinograms. $\chi^2$ maps extend this concept by normalizing residuals against expected noise variance, thereby yielding a statistically grounded measure of local agreement. Whereas image difference maps emphasize absolute deviations, $\chi^2$ maps account for noise and experimental variability and are particularly useful in iterative methodologies in which $\chi^2$ minimization serves as the cost function. \\
As a reconstruction converges toward a correct solution, both difference and $\chi^2$ maps should approach zero values across the field of view, with residuals persisting only in regions where reconstruction remains imperfect. Figure~\ref{fig:comp_ref} illustrates these tools using the Shepp--Logan phantom. In both image and sinogram space, the reconstructed data are compared against their respective references, with the resulting difference and $\chi^2$ maps clearly delineating regions of mismatch. Such visualizations are invaluable for qualitatively diagnosing convergence behavior and guiding \mbox{methodological refinement}.
\begin{figure}[H]
\begin{adjustwidth}{-\extralength}{0cm}
    \centering
    \includegraphics[width=0.99\linewidth]{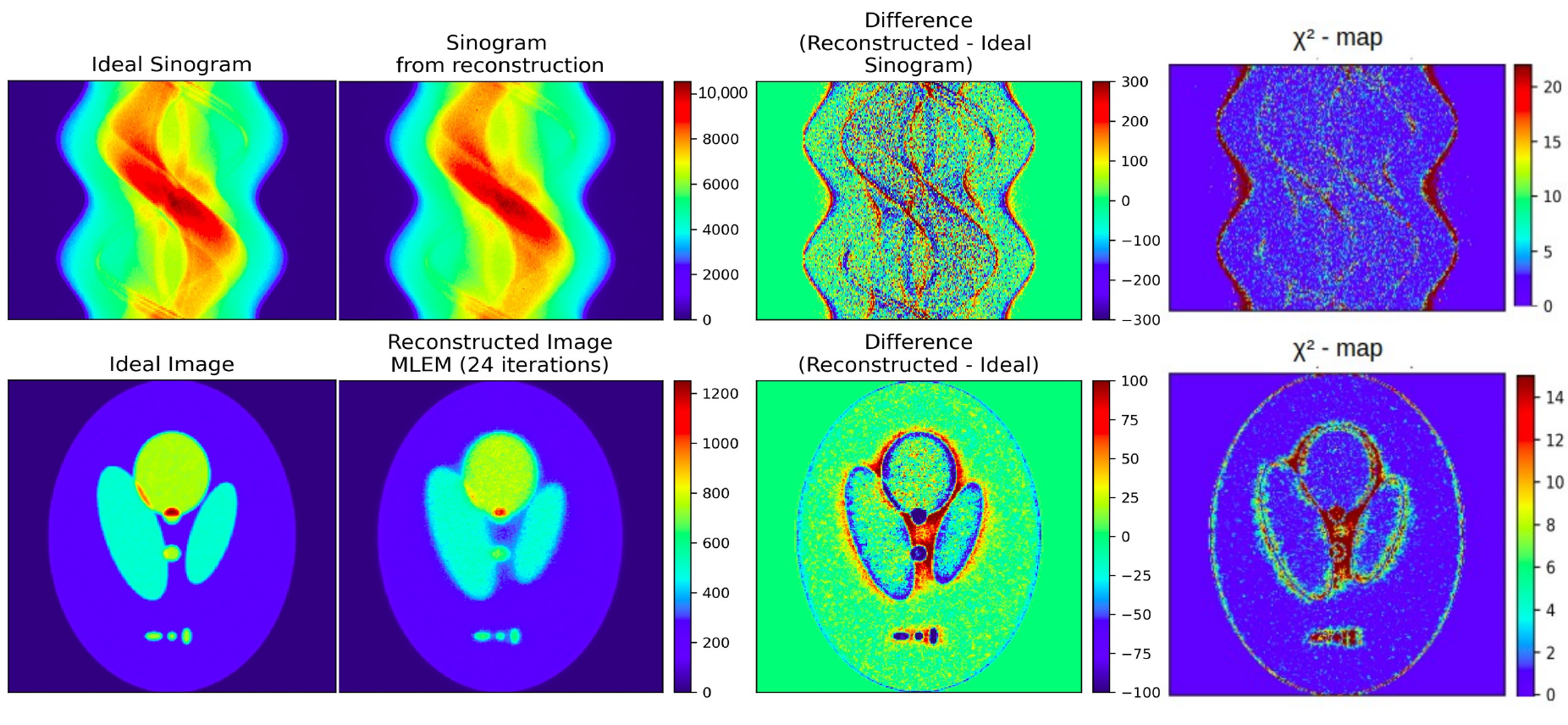}
\end{adjustwidth}
    \caption{
    	Difference 
 and $\chi^2$ map of images and sinograms. In the first row, the sinogram corresponding to an “Ideal Image” of a Shepp--Logan  phantom and the sinogram resulting from an MLEM reconstruction are shown along with the resulting sinogram difference and $\chi^2$ map. In the second row, the “Ideal Image” from a Shepp--Logan phantom, the MLEM reconstructed image, their difference, and the corresponding $\chi^2$ map are shown. }
    \label{fig:comp_ref}

\end{figure}

Formally, let $X$ and $Y$ be two arrays of the same shape representing, respectively, the reconstructed object (image or sinogram) and the reference (typically the ``Ideal'' image or sinogram; for field data, the comparison may be performed in sinogram space). Define the residual (difference) map as
\begin{equation}
R = X - Y
\quad \text{or} \quad
R = |X - Y|.
\label{eq:R_def}
\end{equation}
The signed form visualizes the direction of deviations; the absolute form emphasizes magnitude.\\
To assess the statistical significance of residuals, the pixel-wise (or bin-wise) $\chi^2$ map is computed as
\begin{equation}
\chi^2 = \sum_{i=1}^{N} \frac{(X_{i}-Y_{i})^2}{\sigma_{i}^2}
\label{eq:chi2}
\end{equation}

Spatially correlated clusters of elevated ones indicate modeling inadequacies or systematic reconstruction bias, even when the global $\chi^2$ is acceptable.\\
When $\chi^2$ is the cost function minimized during reconstruction, these maps reveal where the optimization converges effectively and where significant structure remains---information that a single global number cannot convey.

\item \textbf{Structure and Contrast Index (SCI)}

Visual inspection of image difference and $\chi^2$ maps provides valuable qualitative insight into reconstruction deficiencies; however, a scalar diagnostic is required to quantify whether residuals contain organized structure or are dominated by noise, and to enable objective comparison between different reconstruction results. To address this need, we introduce the Structure and Contrast Index (SCI).

The SCI is a diagnostic quantity computed exclusively from a single difference map
\begin{equation}
    R = X - Y
\end{equation}
where X is a reconstructed image (or sinogram) and Y is the corresponding reference, typically the ``Ideal Image'' or ``Ideal Sinogram''. The SCI is not a similarity metric between two images, such as the Structural Similarity Index (SSIM).

\textls[-5]{The SCI is defined as the product of the contrast and structure components that also appear in the SSIM formulation, evaluated on the difference map R. The luminance component of SSIM is explicitly discarded, as it is not meaningful for residual maps whose mean value is close to zero and, in high-fidelity reconstruction regimes, tends to dominate the SSIM value, thereby reducing sensitivity to structured \mbox{residual information.}}

The SCI is introduced to retain sensitivity in high-fidelity reconstruction regimes, where conventional global similarity metrics saturate and lose the ability to discriminate structured residual information. By construction, the SCI quantifies the extent to which the residual map contains coherent, spatially organized structure (e.g., edges, textures, or repeated patterns) as opposed to random, noise-like fluctuations. As reconstruction quality improves, such organized residual structure is progressively eliminated and the SCI approaches zero, whereas elevated values indicate that recoverable information remains in the residuals, pointing to limitations in system modeling, regularization, or reconstruction strategy. Because the SCI is specifically sensitive to structured residual content, its interpretation depends on the nature of the residuals: in regimes dominated by strong, noise-like fluctuations, low values may reflect the masking of coherent structures rather than genuinely high reconstruction fidelity. More generally, as with most quantitative metrics, the SCI should be interpreted in context, alongside complementary metrics and domain-specific considerations, rather than in isolation. This behavior will be illustrated concretely in the benchmarking example discussed in Section~\ref{sec:caseb}.

The SCI can be computed in both image space and sinogram space. For field or clinical data, where a ground-truth image is unavailable, the SCI is most reliably applied in sinogram space, where projection-domain statistics are well-defined.

The values of contrast, structure, and SCI for the difference maps of images and
sinograms shown in Figure~\ref{fig:comp_ref} are given in Table~\ref{tbl:str1}.

\begin{table}[H]

\caption{Contrast and structure components and SCI for difference maps of sinograms and images shown in Figure~\ref{fig:comp_ref}.}
\label{tbl:str1}
\begin{tabularx}{\textwidth}{lCCC}
\toprule
\textbf{Context} & \textbf{Contrast} & \textbf{Structure} & \textbf{SCI} \\
\midrule
Sinograms \\(Difference Map) & 0.07 $\pm$ 0.02 & 0.13 $\pm$ 0.04 & 0.009 $\pm$ $0.004$ \\
Images \\(Difference Map)    & 0.234 $\pm$ 0.001 & 0.242 $\pm$ 0.003 & 0.0566 $\pm$ 0.0007 \\
\bottomrule
\end{tabularx}
\end{table}
\clearpage

\item \textbf{Intensity (gray-value) histogram analysis}

As an additional diagnostic tool, we introduce the concept of intensity histograms analysis of tomographic images and sinograms.
This approach involves analyzing the intensity content (gray value) of an image or sinogram---effectively decomposing it into components that vary in gray value---and comparing the resulting intensity histograms between experimental data and reconstructions.

Given two images $X$ and $Y$, we first normalize them by dividing each image by its mean intensity:
\begin{equation}
X_{\mathrm{norm}} = \frac{X}{\mathrm{mean}(X)}
\quad \text{and} \quad
Y_{\mathrm{norm}} = \frac{Y}{\mathrm{mean}(Y)} .
\label{eq:normalization}
\end{equation}

With this normalization, the total luminosity of the images becomes comparable, since
\begin{equation}
\sum_{i} X_{\mathrm{norm}}^{i}
=
\sum_{i} Y_{\mathrm{norm}}^{i}.
\label{eq:norm_sum}
\end{equation}

Let $S_X$ and $S_Y$ denote their corresponding intensity distributions, represented as one-dimensional histograms using a common set of $NB$ bins. 
All bins have the same width and span the full intensity range required to cover both normalized images.
A direct comparison of the two histograms is now possible. 
Their difference can be quantified in a straightforward way by computing the reduced $\chi^2_{reduced}$ value applied to all histogram bins:
\begin{equation}
\chi^2_{\mathrm{reduced}}
=
\frac{1}{NB}
\sum_{i=1}^{NB}
\left(S_X^{i} - S_Y^{i}\right)^2 .
\label{eq:chi2_red}
\end{equation}

Such comparisons can offer quantitative and visual insight into the fidelity of the reconstruction, particularly with regard to resolution and structural detail.
Reconstructions that exhibit blurring typically show broader or attenuated peaks in their intensity histograms, especially in regions corresponding to high-intensity features such as sharp edges or small hotspots.

Conversely, a well-resolved reconstruction preserves the higher-intensity components present in the original image or sinogram.
Intensity histograms thus complement traditional pixel-domain metrics by probing the intensity domain, where differences in texture, contrast, and resolution become more apparent.

This can be especially revealing for identifying systematic degradation or smoothing introduced by regularization or filtering within the reconstruction pipeline.

The intensity (gray-value) histogram analysis used in this work characterizes the distribution of pixel or bin intensities and should not be confused with spectral or frequency-domain analysis based on spatial Fourier transforms. Histogram analysis probes how reconstruction processes redistribute intensity values and is sensitive to resolution loss, smoothing, and intensity migration effects, whereas spatial-frequency analysis addresses different aspects of image structure.

Figure~\ref{fig:4spectra} provides representative examples of such intensity histograms for the experimental and reconstructed images and sinograms of Figure~\ref{fig:comp_ref}.
As illustrated, deviations in the shape and width of the peaks serve as indicators of the degree to which spatial detail has been preserved or lost during reconstruction.
\begin{figure}[H]
     
    \includegraphics[width=1.0\linewidth]{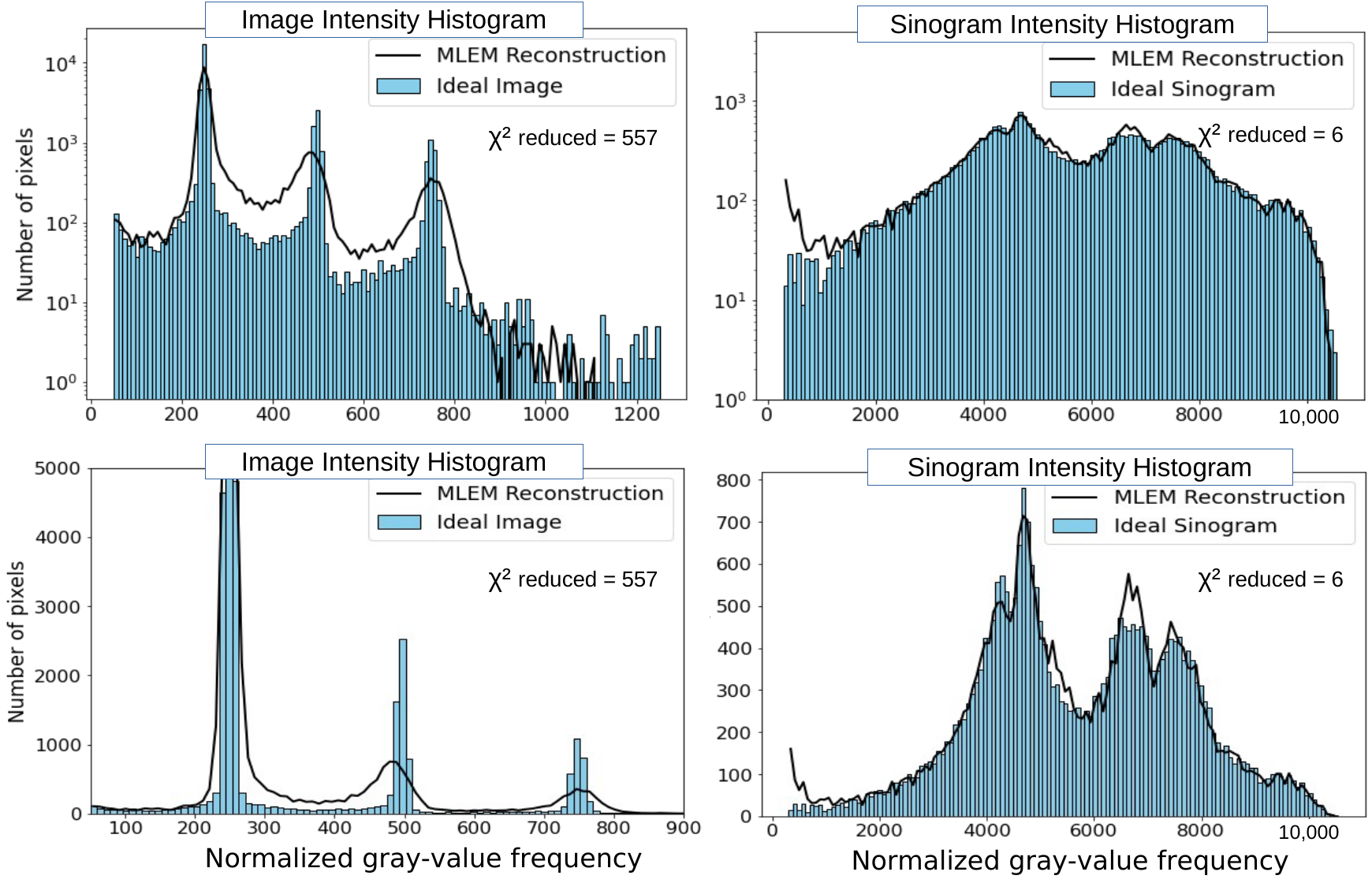}
        \caption{The 
 intensity histograms  (on logarithmic and linear scale) of ``Ideal'' and reconstructed images (on the \textbf{left}) and sinograms (on the \textbf{right}) of Figure \ref{fig:comp_ref} for the case of MLEM after 24 iterations are provided. Linear and logarithmic scales are shown to visualize both dominant and low-frequency components of the gray-value distribution (histogram-based diagnostics quantify differences in gray-value distributions and do not represent spatial-frequency or Fourier-domain analysis
).}
   \label{fig:4spectra}
\end{figure}

\end{enumerate}

\subsubsection{{Region-of-Interest (RoI) Analysis}}\label{sec:roi_def}
A Region of Interest (RoI) is a diagnostic construct that enables localized quantitative evaluation of reconstruction performance within selected subregions of an image~\cite{7422783, 10.1093/scan/nsm006, 4312665, pandey2018automatic, amakusa2018influence, schain2014evaluation, jiang2012regions}.
Conceptually, it functions as a controlled zoom-in on areas of particular diagnostic or structural importance, allowing for quantitative scrutiny of features that may be diluted or concealed in global analyses.

By restricting evaluation to meaningful subregions---such as boundaries, isolated hotspots, or regions containing closely spaced sources---RoIs provide a refined perspective on reconstruction quality and help identify local variations that may not be captured by whole-image metrics.

Although an RoI is defined and processed as an image mask, it is not an additional image type but rather a flexible analytical tool within the evaluation framework.

RoIs are defined on any reference image---e.g., the ``Ideal Image''---and are subsequently applicable to all standardized images introduced in Section~\ref{sec:need_im} (``Source'', ``Detector'', ``Ideal'', and ``Realistic'') that represent the same imaged object or phantom.
The same RoI boundaries are then applied identically to each reconstruction to ensure reproducibility and strict comparability among methodologies.

Because RoIs are defined purely in image space, they have no corresponding sinogram, underscoring that they operate as diagnostic tools rather than data representations.
Each RoI is represented as a binary mask
\[
\Omega \subset \mathbb{Z}^n,
\]
where denotes $\Omega$ the set of discrete pixel or voxel coordinates belonging to the RoI, and $\mathbb{Z}^n$ represents the n-dimensional integer lattice of the image grid (for planar and for volumetric data).
Within each RoI, the quantitative metrics introduced in Sections~\ref{sec:gl_metr} and~\ref{sec:local_tools} can be recomputed to yield localized indicators of fidelity, for example,\vspace{12pt}
\[
\text{SSIM}_{\Omega},\ \text{CNR}_{\Omega},\ \text{NMSE}_{\Omega},\ \chi^{2}_{\Omega}.
\]
These localized metrics quantify reconstruction accuracy exactly where it matters most, avoiding the averaging effects that can obscure deficiencies in global assessments.
The selection of RoIs should be guided by the physical or diagnostic objectives of the study. 

Edge-crossing RoIs evaluate boundary sharpness and partial-volume effects; hotspot RoIs assess localized intensity recovery; and regions containing closely spaced structures probe spatial resolution. Additional RoIs may be defined in uniform or background areas to characterize noise properties or to serve as internal normalization references. When applied systematically, RoI analysis exposes method-dependent differences that remain invisible to global metrics. For instance, two reconstructions may achieve comparable global SSIM but differ markedly in localized SSIM or $\chi^2$ values within critical subregions---highlighting differences in modeling or regularization performance. Thus, RoI analysis serves as an integral “magnifying lens” within the quantitative evaluation toolkit, bridging global metrics and local diagnostics to provide a coherent, multiscale view of \mbox{reconstruction quality.}

Figure~\ref{fig:SL_roi} illustrates the representative RoIs used in this work.
For the Shepp–Logan phantom, three RoIs are shown: (i) an edge-region RoI probing spatial resolution, (ii) a hotspot RoI assessing local intensity recovery, and (iii) a uniform-background RoI quantifying noise stability.

The same masks were applied identically to all reconstructions to ensure reproducibility and facilitate direct comparison between methodologies.

\subsubsection{{Methods of Reconstruction}}\label{sec:methods}
Three distinct reconstruction methods were employed in this study to demonstrate and validate the proposed evaluation framework: the Algebraic Reconstruction Technique (ART)~\cite{Gordon1970, Herman2009},
the Maximum Likelihood Expectation Maximization (MLEM)~\cite{4307558} algorithm, and the recently developed Reconstructed Image from Simulations Ensemble-1 (RISE-1) method~\cite{papanicolas2018novelanalysismethodemission, 9875921, KOUTSANTONIS2019266, 9060020}.

While the first two are well-established iterative frameworks, they differ fundamentally in their optimization logic. ART operates in a deterministic linear way, while MLEM employs a statistical, iterative maximum likelihood criterion to reach the optimal reconstruction. RISE-1 is a novel stochastic, model-based technique that explores a broader parameter space through ensemble sampling.

\begin{itemize}
\item[] \hspace*{-7.5mm}Algebraic Reconstruction Technique (ART)

\end{itemize}

ART is a foundational iterative framework for tomographic imaging, which reformulates the reconstruction problem as a large-scale system of linear equations, seeking to recover the image vector by iteratively projecting current estimates onto the hyperplanes defined by individual projection measurements. Characterized as a row-action method, ART sequentially updates image pixels $f_{j}$ along each ray path to minimize the local residual between the measured data path $S_{i}$ (sinogram) and the forward-projected estimate $R_{i}$, utilizing the projection matrix $P_{ij}$: 
\begin{equation}
f_{j}^{(k+1)} (i) = f_j^{(k)} (i) + \frac{S_i - R_i^{(k)}}{\sum_j P_{ij}} \, P_{ij}
\end{equation}
where (k) denotes the current iteration~\cite{Herman2009}. The deterministic nature of ART, which treats each measurement as an exact linear constraint, provides a robust and computationally efficient framework for addressing ill-posed problems and mitigating streak artifacts characteristic of analytical methods. However, its strict enforcement of data consistency often leads to increased sensitivity to noise and local fluctuations in the reconstructed image.

In the present work, the projection matrix $P_{ij}$ was constructed based on the spatial attenuation distribution of the medium, as determined by GATE Monte Carlo simulations. To facilitate a direct and unbiased comparison with MLEM, the ART algorithm was executed until the $\chi^2$ values exhibited stationarity, serving as a standardized convergence criterion across both iterative methods.

\begin{itemize}
\item[] \hspace*{-7.5mm}Maximum Likelihood Expectation Maximization (MLEM)
\end{itemize}

The MLEM algorithm estimates the activity distribution $f_j$ that maximizes the likelihood of the measured projections $g_i$, given the system response matrix $P_{ij}$. Iterations were performed until convergence, as determined by the stabilization of the global $\chi^2$ between successive reconstructed images.
\begin{equation}
f_{j}^{(k+1)}
=
f_{j}^{(k)}
\frac{1}{\sum_{i} P_{ij}}
\sum_{i} P_{ij}
\frac{g_i}{\sum_{m} P_{im} f_{m}^{(k)}} .
\label{eq:iter_update}
\end{equation}
where $f_{j}^{(k)}$ is the estimated value of voxel $j$ at iteration $k$. 

MLEM reconstructions were generated using synthetic projection data simulated with the GATE (Geant4 Application for Tomographic Emission) toolkit under conditions replicating realistic detector geometry and noise statistics.

\begin{itemize}
\item[] \hspace*{-7.5mm}Reconstructed Image from Simulations Ensemble (RISE-1)
\end{itemize}

The RISE-1 method, part of the RISE suite, is a stochastic, simulation-driven reconstruction technique that generates a set of statistically consistent candidate images by sampling the space of physical and statistical parameters describing the imaging system.
Each ensemble member is obtained by forward-modeling photon transport through the system using the same GATE-based configuration as MLEM, followed by a comparison between simulated and measured projections using a $\chi^2$ cost function.
The final reconstructed image is calculated as the expectation value over the accepted ensemble members, yielding an inherently regularized solution that suppresses noise amplification and reduces dependence on initial conditions.

Both reconstruction methods were applied to the standardized phantom data introduced earlier.
For MLEM, reconstructions were generated at successive iteration counts to analyze convergence behavior and to demonstrate the sensitivity of the proposed metrics and diagnostic tools to subtle image improvements.
For RISE-1, the reconstruction was performed using the same projection data and imaging geometry, enabling direct quantitative comparison with the MLEM results.

\section{Results}\label{sec:results2}
In this section, we demonstrate the utility and versatility of the standardized reference images, simulated sinograms, and evaluation metrics introduced in the previous section. Designed for broad applicability, these tools enable rigorous, quantitative comparison across diverse reconstruction methodologies and imaging modalities.

To illustrate their relevance, two representative case studies are presented.
In Case A, we analyze the behavior and convergence of the MLEM algorithm~\cite{Hwang_2006}, using both conventional and newly introduced metrics to reveal subtle changes in image quality and reconstruction performance. The study of the well-known MLEM illustrates the value of the several diagnostic tools and metrics without adding the complexity of an \mbox{unfamiliar algorithm.}

In Case B, we evaluate the novel RISE-1 algorithm, directly benchmarking it against MLEM and ART using the same identical data, phantoms, and reference images. This comparison highlights the ability of the proposed method to sensitively detect differences in reconstruction quality---both globally and within selected Regions of Interest (RoIs)---providing insights beyond those offered by traditional single-value measures. A comparative assessment of RISE-1, MLEM, and ART is conducted to demonstrate the methodological framework, which is designed to support transparent and reproducible comparisons between established algorithms and novel or less familiar approaches.

\textls[-15]{As noted in Section~\ref{sec:gl_metr}, our analysis considered a broader family of global scalar and geometric comparison metrics---including information divergences and optimal-transport distances---that can offer enhanced sensitivity in high-fidelity regimes. For the sake of methodological consistency and clarity in the comparative studies that follow, we report exclusively the $\chi^2$ statistic in this section, even though more advanced metrics were examined in parallel.}

\subsection{{CASE A: Exploring Convergence and Properties of MLEM}}\label{sec:caseA}
\begin{figure}[H]
    \centering
    \begin{adjustwidth}{-\extralength}{0cm}
    \includegraphics[width=1.0\linewidth]{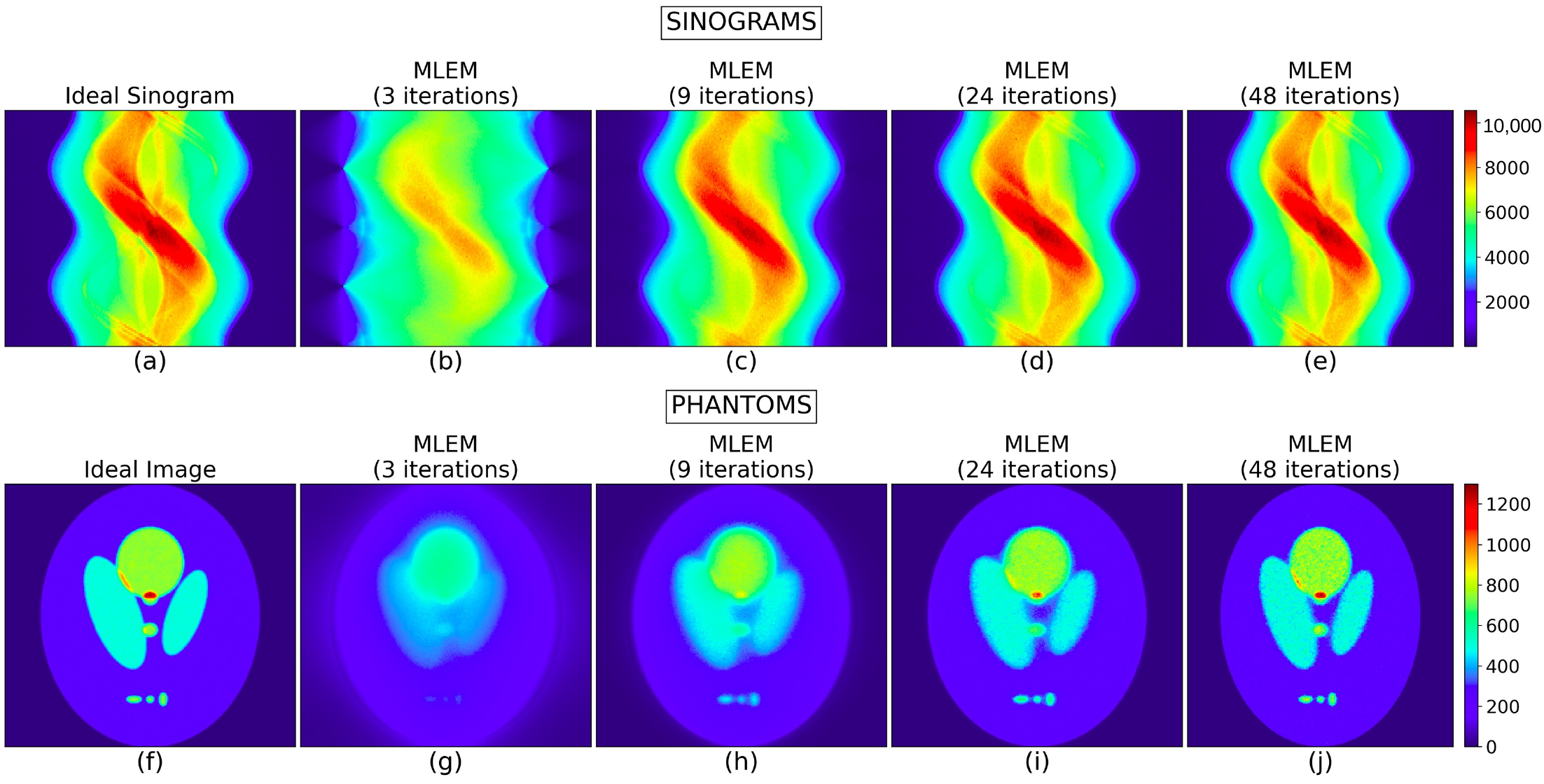}
    \end{adjustwidth}
    \caption{The 
 top row depicts the ``Ideal Sinogram'' (\textbf{a}) and those resulting from different stages of MLEM (3, 9, 24, and 48 iterations) (\textbf{b}–\textbf{e}). The bottom row shows the ``Ideal Image'' (\textbf{f}) and the corresponding MLEM reconstructions (\textbf{g}–\textbf{j}).}
    \label{fig:entire_ph}
\end{figure}

\begin{figure}[H]
     
 
    \includegraphics[width=0.85\linewidth]{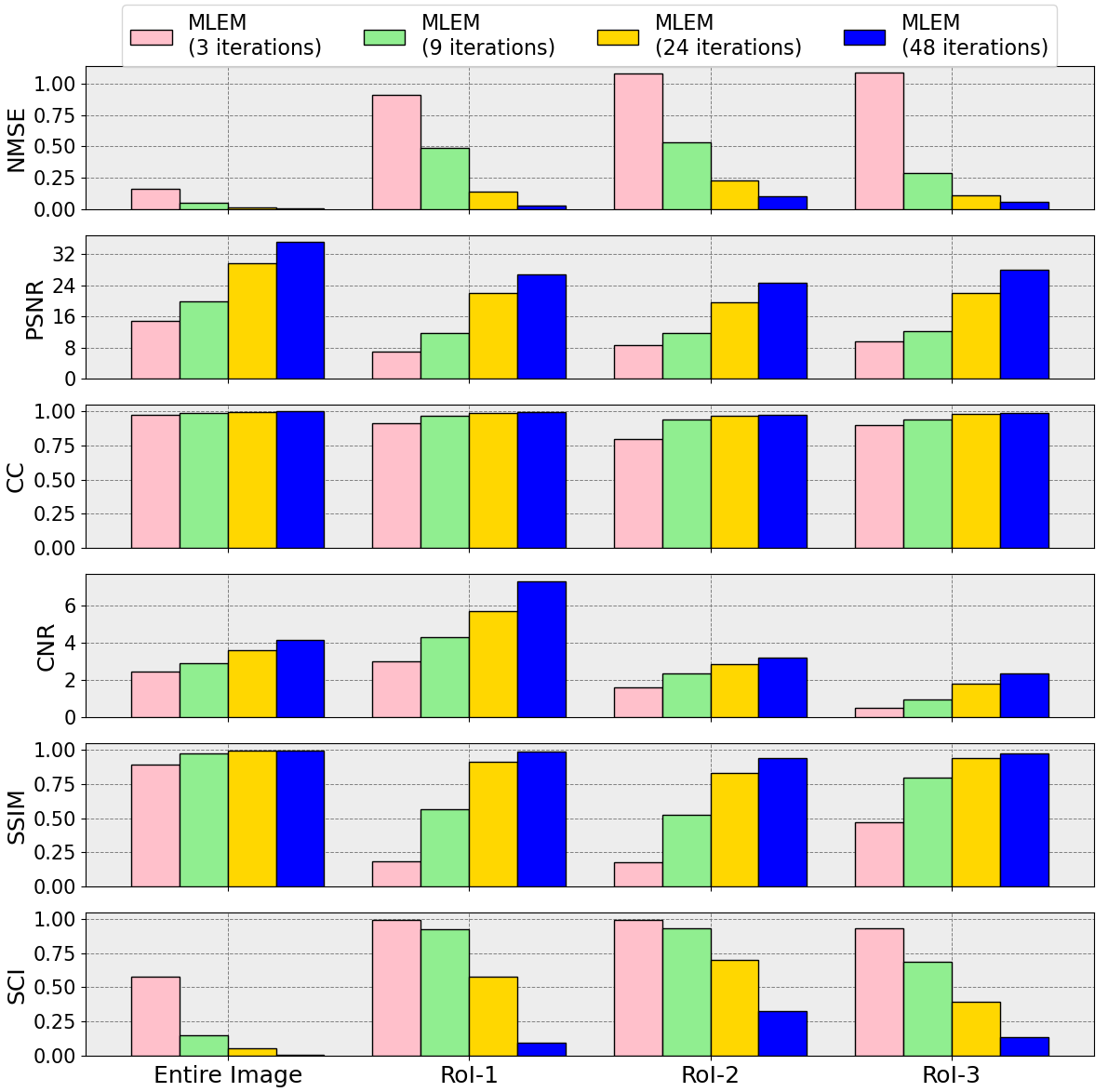} 
    \caption{A selection of metrics (NMSE, PSNR, CC, SSIM, CNR, and SCI) are presented for each one of the  RoIs and the entire image, and for four stages of MLEM evolution. The SCI computed from the difference map between reconstructed and reference images. SSIM approaches unity at early reconstruction stages, whereas the SCI continues to evolve, indicating sensitivity to residual structure beyond global similarity.}
    \label{fig:bar_mlem}
\end{figure}

In this first case, we chose to showcase the method using a 2-D software phantom (Shepp--Logan: Table~\ref{tab:shepp_logan} and Figure~\ref{fig:SL_roi}). The sinograms were generated using \mbox{128 equally spaced projections} by simulating the emission and propagation of the photons with GEANT4/GATE using an “Ideal Collimator”. All simulated experiments were performed using controlled random seeds and identical instrument parameters to ensure reproducibility. Data generation and pre-processing settings were kept constant across all evaluations. To assess robustness, all reconstructed images were generated with comparable count statistics consistent with the Poisson processes governing $\gamma$-ray emission. Under these conditions, the MLEM algorithm demonstrated stable performance with only minor variations. For the case of Shepp--Logan a number of  MLEM reconstructions were produced, each resulting from a different number of iterations. Four of them are shown in Figure \ref{fig:entire_ph}, alongside the “Ideal Image'' and their corresponding sinograms. Metric uncertainties were estimated using the Jackknife method~\cite{busing1999delete, Quenouille1956, tukey1958}. The global metrics values are illustrated in the first column of Figure
~\ref{fig:bar_mlem}. Their numerical values along with the metrics in the sinogram domain and a comparison of MLEM (48 iterations) with ``Source Image'' instead of ``Ideal'' are provided in Appendix~\ref{app:gl_mlem}.
\subsubsection{{$\chi^2$ and Difference Maps for Sinograms and Images}}\label{sec:dif}
\begin{figure}[H]
\begin{adjustwidth}{-\extralength}{0cm}

    \centering
    \includegraphics[width=0.9\linewidth]{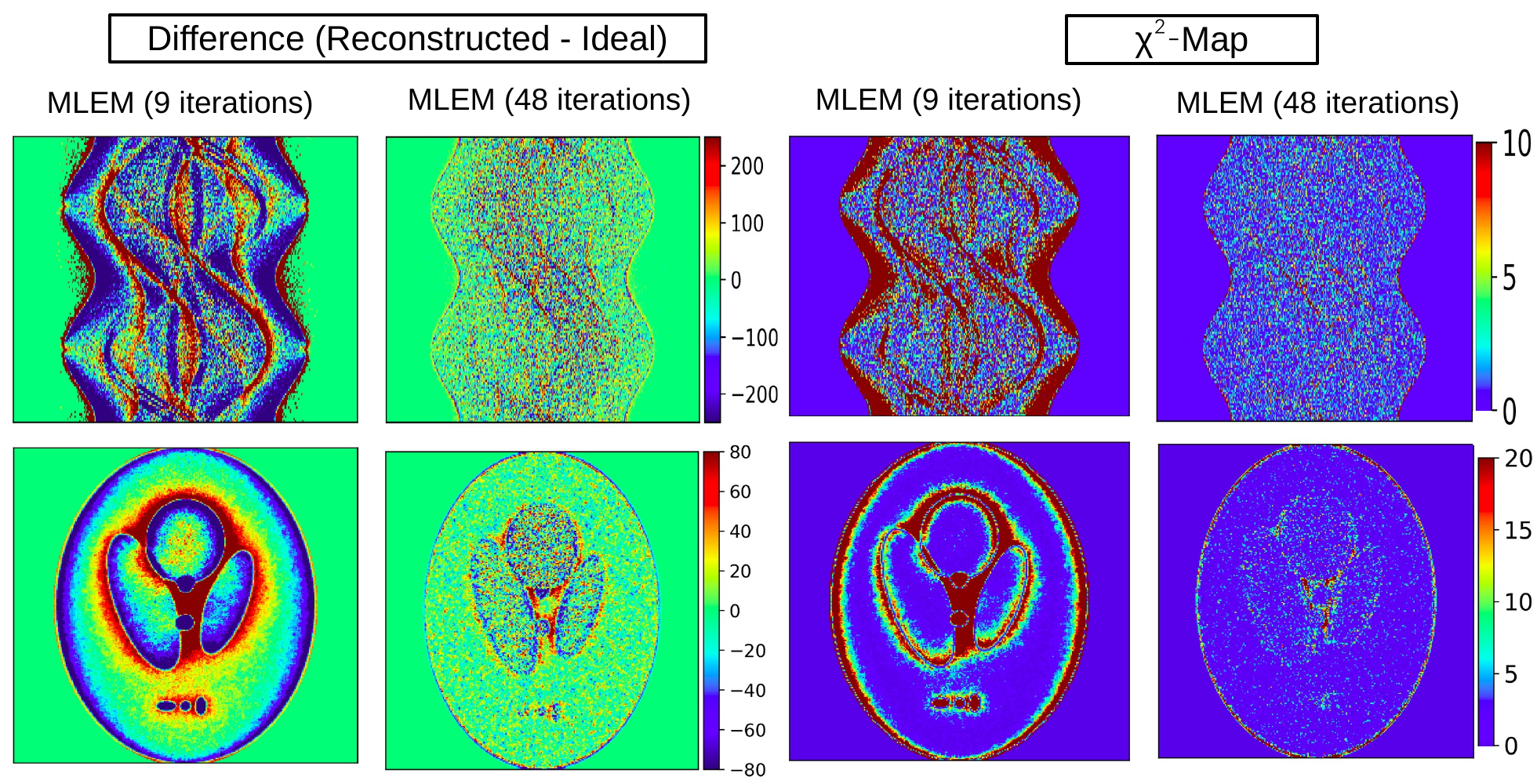}
\end{adjustwidth}
    \caption{
    	 The first row provides the 
    difference and $\chi^2$ maps for the sinogram and the second one the 
    difference and $\chi^2$ maps for the images. The difference and $\chi^2$ maps reveal areas of the images that are not well-represented by the reconstruction.}
    \label{fig:9_48iter}
\end{figure}

As discussed in Section~\ref{sec:local_tools}, difference maps and $\chi^2$ maps provide further insight into the quality of reconstructions, highlighting both absolute deviations and statistically normalized ones, respectively. In this case study, we apply these diagnostics to evaluate MLEM reconstructions at two stages, after 9 and 48 iterations, as illustrated in Figure~\ref{fig:9_48iter}. Notably, in the fully converged case, the $\chi^2$ map of the sinogram ($\chi^2_{reduced}$ = 2) shows minimal residual structure, indicating convergence, indicating that minimal additional information can be extracted from the data. By contrast, the maps obtained after nine iterations ($\chi^2_{reduced}$ = 12) reveal clear residual patterns, underscoring that the reconstruction has not yet converged and it has not captured the available information. The difference maps (Figure~\ref{fig:9_48iter}), in contrast to the $\chi^2$ maps, also provide the sign of the discrepancy in each area, allowing us to identify which regions in the image and sinogram are overestimated or underestimated. In the difference map of the MLEM image after nine iterations, we notice that the background around the hotspots is significantly overestimated, while the boundaries of the hotspots are underestimated---an effect that is greatly mitigated after 48 iterations.

\subsubsection{{Structure and Contrast Index (SCI) of Images and Sinograms} } \label{sec:sci}
\begin{figure}[H]
    \centering
    \begin{adjustwidth}{-\extralength}{0cm}
    \includegraphics[width=1.0\linewidth]{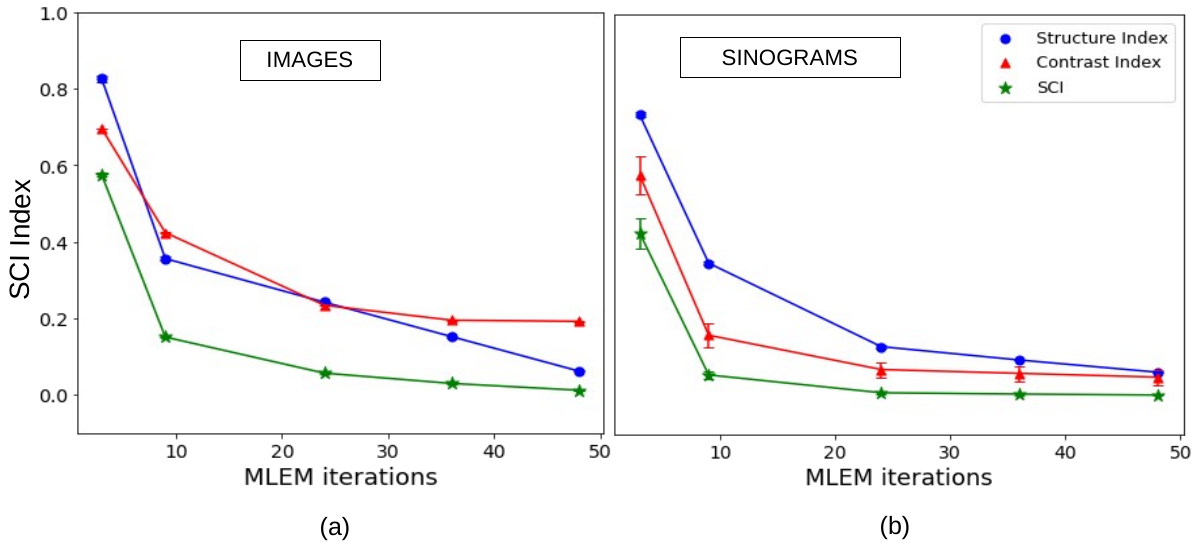}
    \end{adjustwidth}
    \caption{Convergence of the SCI. The structure, contrast, and SCI are computed from the difference map between reconstructed (four MLEM stages) and reference images in (\textbf{a}) and sinograms in (\textbf{b}).}
    \label{fig:sci_im_sino}
\end{figure}

The Structure and Contrast Index (SCI) is the metric we introduced to quantify the magnitude of the discrepancy seen in the difference maps of sinograms and images.  
The values of the SCI, along with the individual components of Luminance Contrast and Structure, for image and sinogram difference maps for all the stages of MLEM are given in Table \ref{tbl:ssim} and illustrated in Figure~\ref{fig:sci_im_sino}. As expected, the SCI of the sinogram difference map for MLEM (three iterations) is higher compared to the case of MLEM at full convergence. As the algorithm converges, the difference becomes dominated by statistical fluctuation and noise, leading to a lower SCI for both image and sinogram difference maps, indicating that little additional information can be obtained from the experimental data.

While global similarity metrics such as SSIM rapidly approach unity after only a few reconstruction iterations, 
 the SCI---computed on difference maps---continues to decrease systematically, reflecting the progressive elimination of structured residuals. This demonstrates that the SCI provides complementary information rather than duplicating the role of SSIM, particularly in the high-fidelity regime where global similarity metrics saturate (Table~\ref{tbl:ssim} and Figure~\ref{fig:bar_mlem}).

\begin{table}[H]
\caption{Luminance
, contrast, structure, and SCI values of sinogram and image difference maps (reconstructed---``Ideal'') using MLEM at different iteration stages.}
\label{tbl:ssim} 
\begin{adjustwidth}{-\extralength}{0cm}
\renewcommand{\arraystretch}{1.2}
\begin{tabularx}{\fulllength}{LccCC}
\toprule
 \textbf{Metric on Difference Map} 
& \textbf{MLEM (3 Iterations)} 
&  \textbf{MLEM (9 Iterations)} 
&  \textbf{MLEM (24 Iterations)} 
&  \textbf{MLEM (48 Iterations)} \\
\midrule
Luminance (sinogram) & 0.000 ± 0.006  & 0.000 ± 0.002   & 0.000 ± 0.002   & 0.000 ± 0.002 \\
 Luminance (image)   & 0.000 ± 0.002 & 0.0000 ± 0.0002 & 0.0000 ± 0.0004 & 0.0000 ± 0.0002 \\
\hline
Contrast (sinogram) & 0.58 ± 0.05    & 0.16 ± 0.03     & 0.07 ± 0.02     & 0.05 ± 0.02 \\
Contrast (image)    & 0.696 ± 0.001  & 0.424 ± 0.001   & 0.234 ± 0.001   & 0.192 ± 0.001 \\
\hline
Structure (sinogram) & 0.738 ± 0.007  & 0.349 ± 0.005   & 0.130 ± 0.004    & 0.063 ± 0.001 \\
Structure (image)   & 0.826 ± 0.007  & 0.356 ± 0.005   & 0.242 ± 0.003  & 0.062 ± 0.002 \\
\hline
SCI (sinogram) & 0.43 ± 0.04  & 0.06 ± 0.01   & 0.009 ± 0.003    & 0.003 ± 0.001 \\
SCI (image)  & 0.575 ± 0.005  & 0.151 ± 0.002   & 0.057 ± 0.001  & 0.0119 ± 0.0004 \\
\noalign{\hrule height 1pt}
\end{tabularx}
\end{adjustwidth}
\end{table}

\vspace{-6pt}

\subsubsection{{Intensity (Gray-Value) Histogram Analysis of Images and Sinograms}}\label{sec:spectrum}
We have introduced the intensity (gray-value) histogram analysis of tomographic images and their corresponding sinograms as an additional diagnostic tool. 

\begin{figure}[H]
    \centering
    \begin{adjustwidth}{-\extralength}{0cm}
    \includegraphics[width=1.0\linewidth]{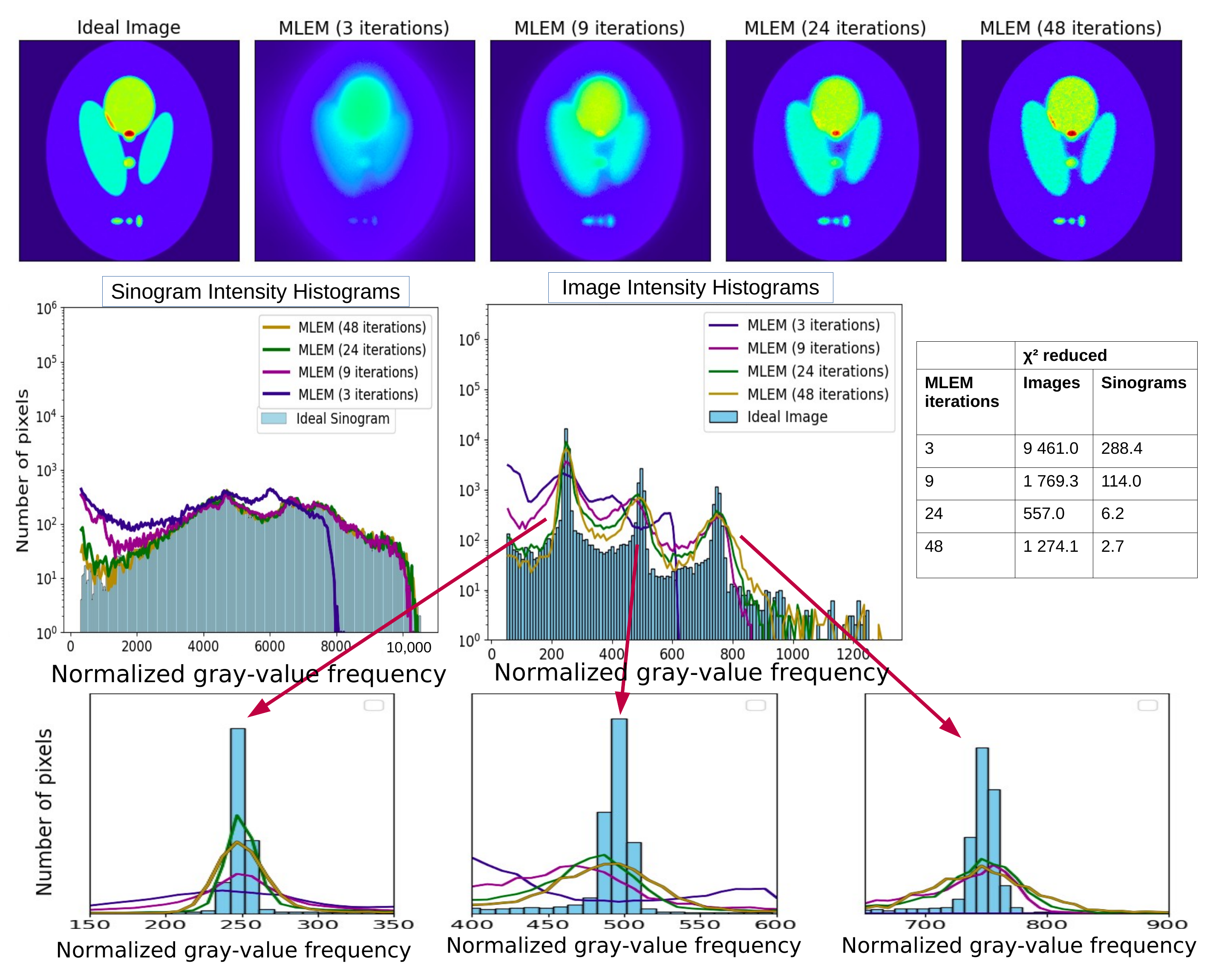}
    \end{adjustwidth}
    \caption{The 
 top row of the figure displays the ``Ideal Image'' alongside the four MLEM reconstructions. In the second row, the intensity histograms of the ``Ideal Image'' and each of the four reconstructions are illustrated on the left. On the right, the intensity histograms of the ``Ideal Sinogram'' and those resulting from MLEM are presented. Linear and logarithmic scales are shown to visualize both dominant and low-frequency components of the gray-value distribution (histogram-based diagnostics quantify differences in gray-value distributions and do not represent spatial-frequency or Fourier-domain analysis
).}
    \label{fig:ola_mlem}
\end{figure}

In Figure~\ref{fig:ola_mlem}, we study the image and sinogram intensity histograms for several MLEM stages (3, 9, 24, and 48 iterations).

\begin{figure}[H]
    \centering
    \begin{adjustwidth}{-\extralength}{0cm}
    \includegraphics[width=1.0\linewidth]{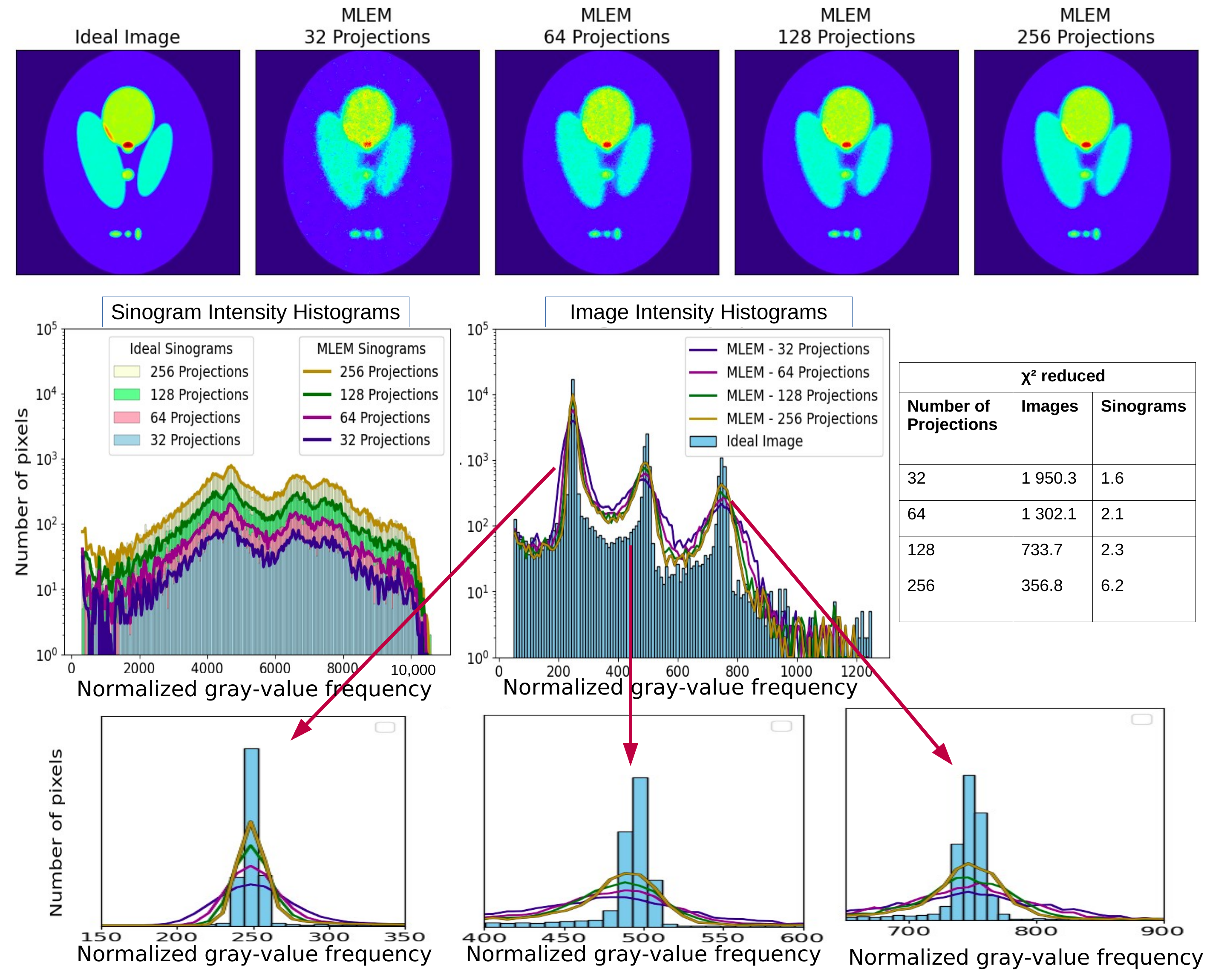}
    \end{adjustwidth}
    \caption{The 
 top row of the figure displays the ``Ideal Image'' alongside the three reconstructions from MLEM at full convergence and with 32, 64, and 256 projections. In the second row, the intensity histograms of the ``Ideal Image'' and each of the three reconstructions are illustrated on the left. On the right, the intensity histograms of the ``Ideal Sinogram'' and those resulting from 32, 64, and 256 projections are presented. Linear and logarithmic scales are shown to visualize both dominant and low-frequency components of the gray-value distribution (histogram-based diagnostics quantify differences in gray-value distributions and do not represent spatial-frequency or Fourier-\mbox{domain analysis}
).}
    \label{fig:ola_iter}
\end{figure}

In addition we show the utility of this tool investigating how the number of angular projections used in reconstruction affects image quality (Figure~\ref{fig:ola_iter}). A Shepp--Logan phantom, simulated using an ``Ideal Collimator,'' serves as the ground truth. Reconstructions were performed using sinograms with 32, 64, and 256 uniformly distributed \mbox{planar projections.} 

In Figure \ref{fig:ola_iter}, we present the intensity histograms of the ``Ideal Image'' and those of the reconstructed images for each projection set. As the number of projections increases, the intensity histogram peaks in the reconstructed images become narrower, progressively converging toward the intensity histogram of the ``Ideal Image''. This demonstrates and quantifies that increased angular sampling enhances the reconstruction’s ability to recover accurate intensity distributions, particularly around localized high-intensity regions (e.g., ``hotspots''), resulting in higher resolution.

\begin{figure}[H]
    
    \begin{adjustwidth}{-\extralength}{0cm}\centering
    \includegraphics[width=0.95\linewidth]{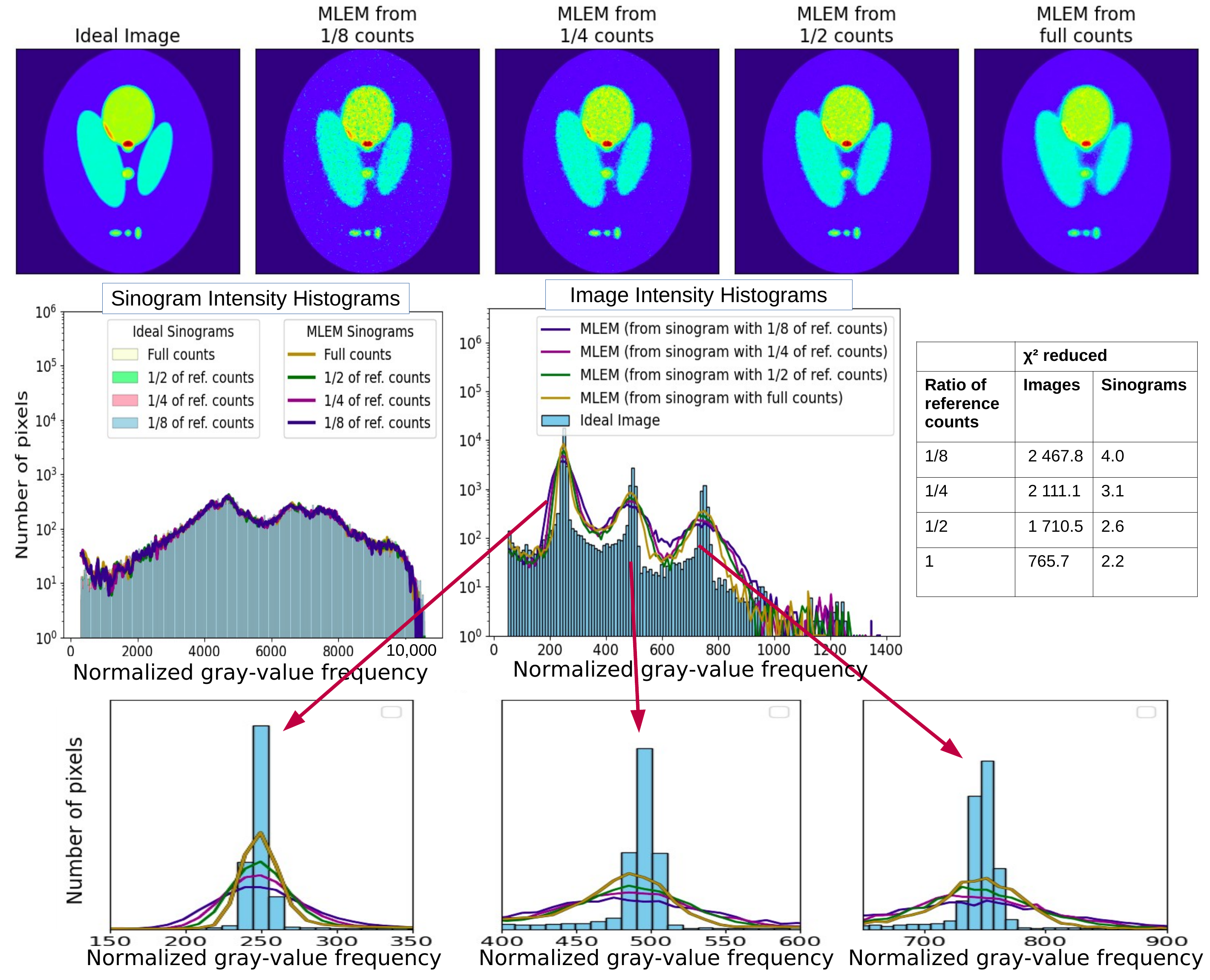}
    \end{adjustwidth}
    \caption{The 
 top row of the figure displays the ``Ideal Image'' alongside the four reconstructions from MLEM at full convergence, utilizing sinograms with different level of statistics (1/8, 1/4, 1/2, and full counts). In the second row, the intensity histograms of the ``Ideal Image'' and each of the four reconstructions are illustrated on the left. On the right, the intensity histograms of the ``Ideal Sinograms'' with the four cases  are presented. Linear and logarithmic scales are shown to visualize both dominant and low-frequency components of the gray-value distribution (histogram-based diagnostics quantify differences in gray-value distributions and do not represent spatial-frequency or Fourier-\mbox{domain analysis}
).}
    \label{fig:ola_statistics}
\end{figure}

Interestingly, Figure \ref{fig:ola_iter}, which displays the sinogram intensity histograms, shows that the overall shape of the intensity histogram of the sinograms remains rather insensitive to the number of projections. This indicates that while the information content of the sinogram may appear similar, the resolution and fidelity of the final reconstructed image are significantly influenced by the density of angular sampling. The intensity histogram of the image capture with greater sensitivity than sinogram-based metrics alone.
In the next example (Figure~\ref{fig:ola_statistics}), 
we present the image intensity histograms and sinogram intensity histograms obtained from different levels of statistics. Artifacts during reconstruction---because of statistical fluctuation on the sinograms that allow for the migration of image intensity to neighboring pixels---contribute to the distortion of the intensity histogram. Higher levels of statistics lead to sharper peaks on the intensity histogram.

The two examples shown demonstrate that the intensity histogram analysis is a valuable diagnostic tool for assessing reconstruction quality. In addition, the intensity histogram of the sinograms shows that no area with different numbers of counts reaching the detector is missed or underestimated. It is important to analyze both the sinograms and the images, as agreement in the sinogram's intensity histogram does not guarantee the agreement in the image’s intensity histogram. 

\subsubsection{{Region-of-Interest (RoI) Analysis: Application of New Metrics and Images}}\label{sec:rois}
A limitation of global evaluation metrics---such as $\chi^2$ or SSIM---is that they dilute the effect of localized discrepancies across the full image or sinogram. As a result, small but diagnostically critical deviations (e.g., missed hotspots; poor edge delineation) may go undetected. To remedy this, as discussed in Section~\ref{sec:roi_def}, we make use of the well-known technique  of Region-of-Interest (RoI) analysis~\cite{7422783, 10.1093/scan/nsm006, 4312665, pandey2018automatic, amakusa2018influence, schain2014evaluation, jiang2012regions} in conjunction with the newly introduced images and metrics.
We defined in Section~\ref{sec:roi_def} (see Figure~\ref{fig:SL_roi}) three representative RoIs in the Shepp--Logan phantom to target specific reconstruction challenges:
For each of these RoIs, we compute localized $\chi^2$ values, intensity histograms analysis, and SSIMs of the reconstructed and reference images or sinograms. These localized metrics offer a powerful complement to global scores, as they reveal details that may go unnoticed otherwise.

In Appendix~\ref{app:roi} we present the RoIs of the ``Ideal Image'' alongside those from the four MLEM reconstructions and a table with all the metrics values for each RoI. The corresponding bar plots for each RoI, along with the ones for the entire image, are presented in Figure~\ref{fig:bar_mlem}. The metrics show greater sensitivity in RoIs than the entire image, as expected. For example, SSIM values for RoIs are significantly lower than those computed for the entire image, emphasizing the importance of localized evaluation in identifying performance gaps. Additionally, for the newly introduced metrics SCI, the values for RoIs showed greater variation across MLEM iterations, and substantially higher values, about an order of magnitude, indicative of its sensitivity, further underscoring the utility of RoI analysis in tracking convergence and identifying areas for improvement.

By leveraging RoI analysis in conjunction with the new metrics, we validate the sensitivity and interpretability of the proposed metrics and tools, demonstrating their ability to characterize local image fidelity and optimize reconstruction methods. 
\subsection{{CASE B: Benchmarking a New Reconstruction Method}} \label{sec:caseb}
In this second demonstration case, we apply the new methodology to evaluate and benchmark the performance of RISE-1 (Section \ref{sec:methods}), a novel algorithmic reconstruction method. Using the introduced images and quantitative metrics, we compare the performance of the novel RISE-1 against the established MLEM and ART methods. 

The proposed methodology is reconstruction-method agnostic: it does not depend on the internal formulation of the algorithm, whether proprietary, unpublished, or otherwise inaccessible. Instead, it evaluates performance solely on the basis of the reconstructed output, ensuring a fair, consistent, and broadly applicable benchmarking framework.

To demonstrate this benchmarking process, we selected a particularly challenging test case: the reconstruction of a Shepp--Logan phantom with significant background activity. The specific activity ratio across regions (background:side ellipses:central ellipse and hotspots) is set to (20:5:10), according to Table~\ref{tab:shepp_logan}. In this case study, the phantom configuration is intentionally more challenging than in the MLEM convergence study, with background activity increased by a factor of four. This elevated background reduces contrast and makes the reconstruction problem intrinsically more difficult. The ``Ideal Sinograms'' were generated using the ``Ideal Collimator'' model, ensuring a high-quality reference dataset for evaluation.

\begin{figure}[H]
     
\begin{adjustwidth}{-\extralength}{0cm}
\centering 

    \includegraphics[width=0.9\linewidth]{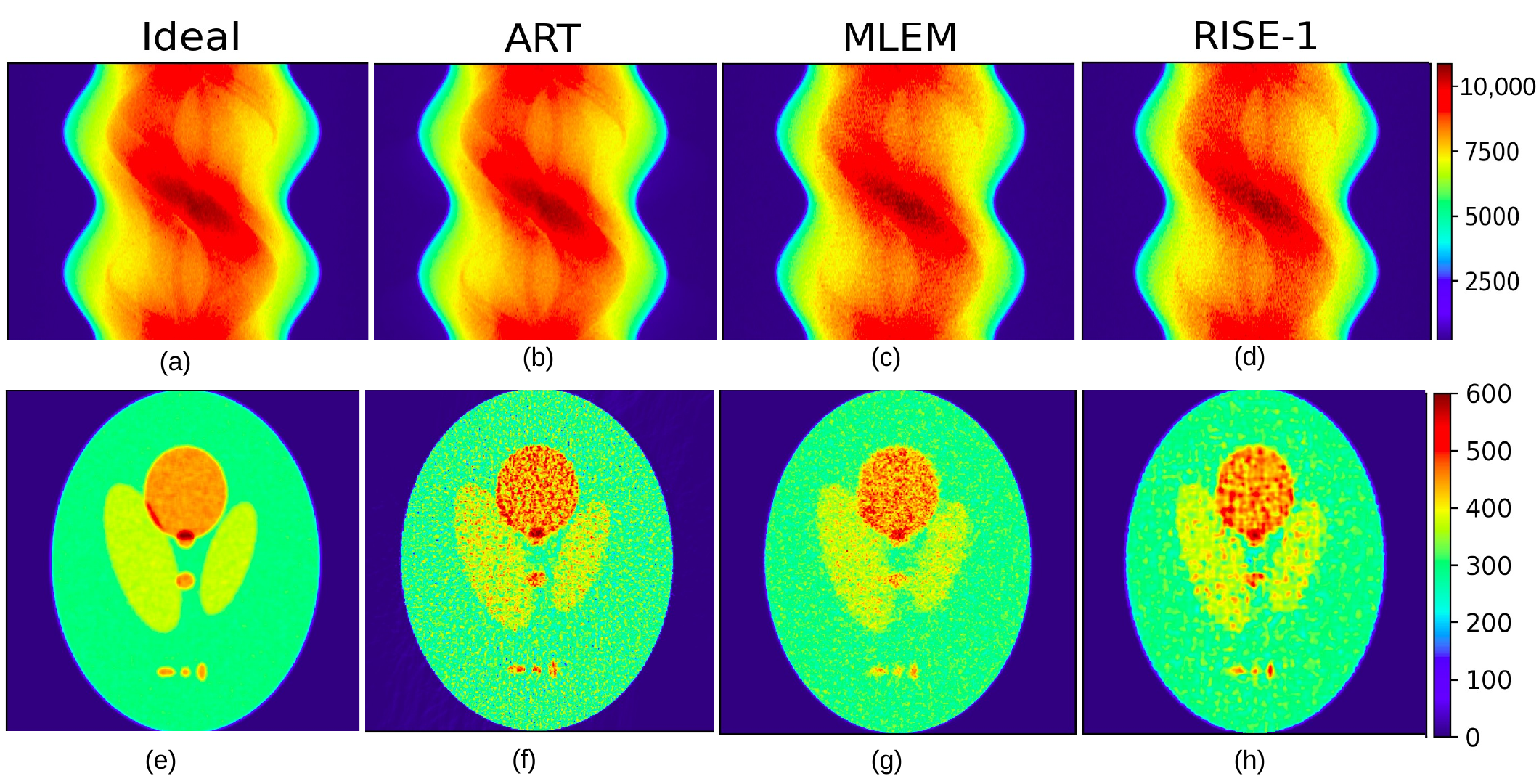}
\end{adjustwidth}
  \caption{Top 
 row: (\textbf{a}) ``Ideal Sinogram'' and sinograms obtained using three reconstruction methods: (\textbf{b}) ART, (\textbf{c}) MLEM, and (\textbf{d}) RISE-1. Bottom row: (\textbf{e}) “Ideal Image” and corresponding reconstructions: (\textbf{f}) ART, (\textbf{g}) MLEM, and (\textbf{h}) RISE-1.}
    \label{fig:rise}
\end{figure}

Figure~\ref{fig:rise} presents the ART, MLEM, and RISE-1 reconstructions alongside the “Ideal Image” and their corresponding sinograms. A visual inspection suggests broadly comparable performance among the three algorithms. All methods successfully capture the overall structure and reproduce the hotspots, providing a faithful representation of the phantom.

\begin{figure}[H]
\begin{adjustwidth}{-\extralength}{0cm}

    \centering
    \includegraphics[width=0.9\linewidth]{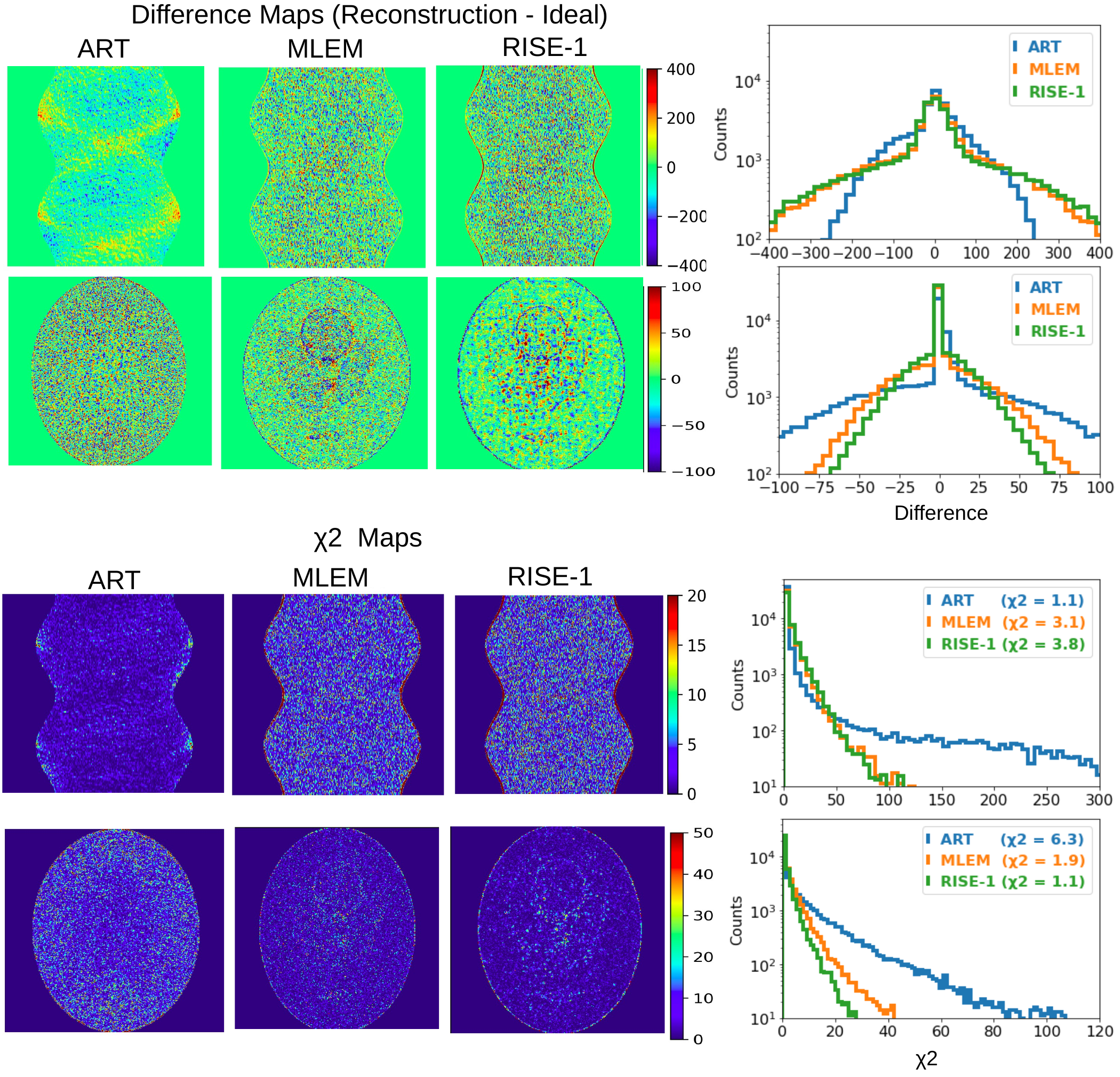}
\end{adjustwidth}
    \caption{
    	The top-left and bottom-left panels show the difference maps (reconstruction---``Ideal'') and $\chi^2$ maps for images and sinograms, for the three reconstruction methods (ART, MLEM, and RISE-1). The right panels present the corresponding residual intensities and $\chi^2$ histograms. The $\chi^2_{reduced}$ values for both images and sinograms are reported in the bottom $\chi^2$ histograms for each reconstruction method.}
    \label{fig:rise_im_sino}
\end{figure}

The examination of the metrics associated with the overall reconstruction (Table~\ref{tab:metrics}) indicates that RISE-1 and MLEM perform comparably and consistently outperform ART. Particularly informative are the difference maps (Figure~\ref{fig:rise_im_sino}), which reveal that the improved $\chi^2$ of ART in representing the sinogram is achieved at the expense of introducing pronounced pixel-to-pixel fluctuations in the reconstructed image.

\begin{figure}[H]
     
\begin{adjustwidth}{-\extralength}{0cm}
\centering 

    \includegraphics[width=1.0\linewidth]{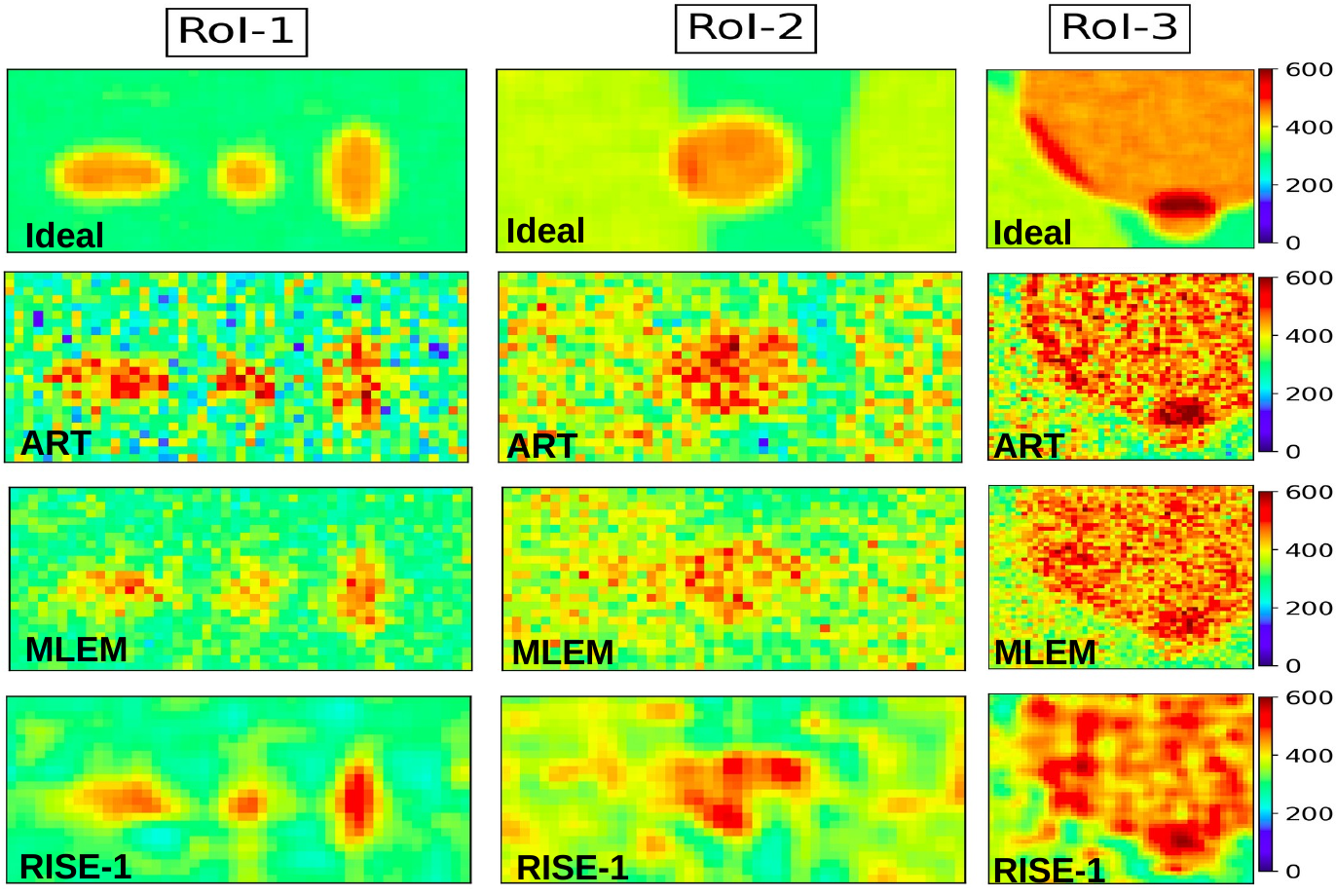}
\end{adjustwidth}
    \caption{ The panels show three RoIs from the phantom. The top row presents the ``Ideal Image'', followed by reconstructions obtained with ART, MLEM, and RISE-1 in the subsequent rows. The images are normalized with reference to ``Ideal Image''.}
    \label{fig:rise_roi}
\end{figure}

\begin{figure}[H]
\begin{adjustwidth}{-\extralength}{0cm}

    \centering
    \includegraphics[width=\linewidth, height=0.8\linewidth]{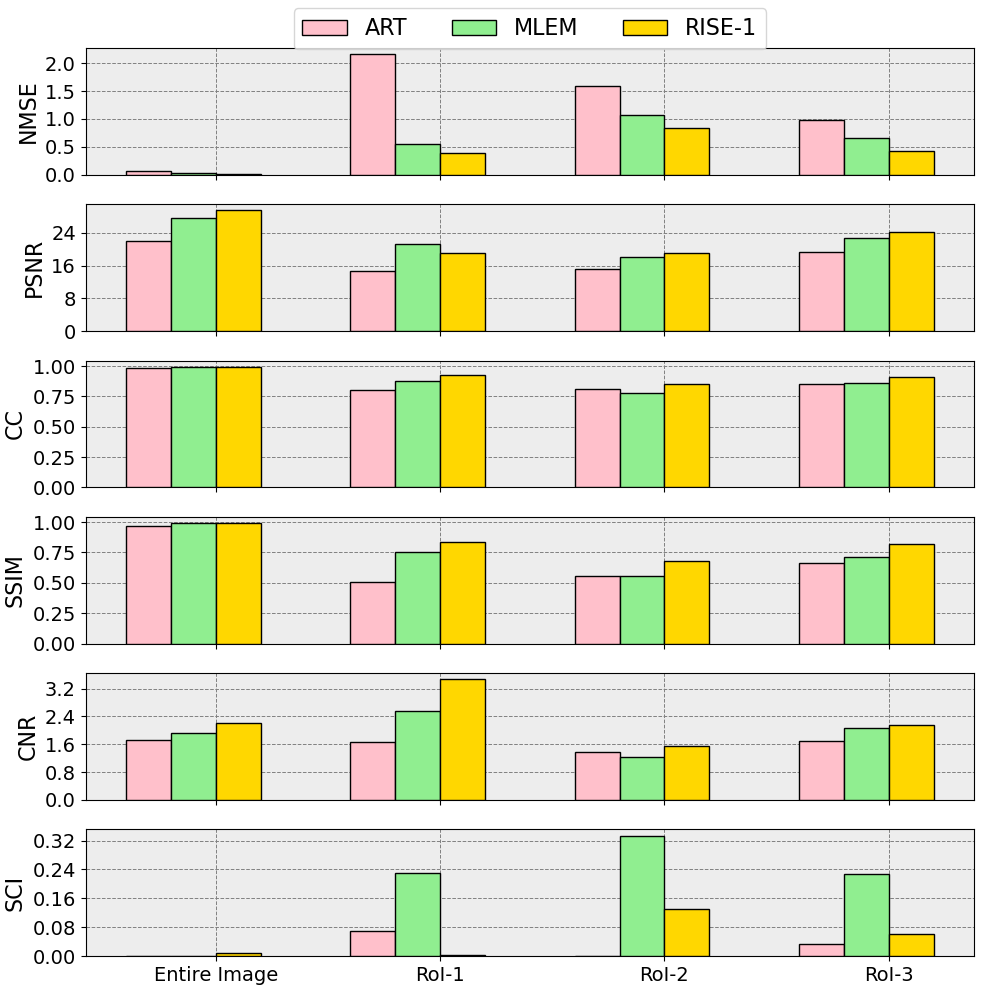}
\end{adjustwidth}
    \caption{The metrics NMSE, PSNR, CC, CNR, SSIM, and SCI are presented for each one of the RoIs
and the entire image, across the three reconstruction techniques: ART, MLEM, and RISE-1. The SCI
computed from the difference map between reconstructed and reference area. The reconstructions
shown correspond to the high-background phantom configuration described in Section~\ref{sec:caseb} and
Table~\ref{tab:shepp_logan}, which is substantially more challenging than the phantom used in the MLEM convergence
study (Figure~\ref{fig:bar_mlem}).}
    \label{fig:rise_bar}
\end{figure}

To further discriminate between the two high-performing methods (MLEM and RISE-1), we examine the RoIs shown in Figure~\ref{fig:rise_roi}, together with the corresponding metric values reported in Table~\ref{tab:metrics} and the associated plots in Figure~\ref{fig:rise_bar}. Across all RoIs, the small hotspots are more accurately reproduced by RISE-1, as reflected in higher values of CC, SSIM, and CNR and lower NMSE and SCI values. Notably, the newly introduced SCI exhibits the strongest discriminative power between these near-convergent methods.

ART, which is found to underperform relative to MLEM and RISE-1 according to global metrics, attains even lower SCI values, which may appear counterintuitive. However, this behavior is fully consistent with the interpretation of the SCI in fluctuation-dominated regimes. The pronounced pixel-to-pixel fluctuations characteristic of ART reconstructions---clearly evident in Figure~\ref{fig:rise} and further reflected in the difference maps (Figure~\ref{fig:rise_im_sino})---effectively mask any remaining coherent residual structure. As a result, the SCI, being specifically sensitive to structured residual content, returns low values not because the reconstruction is superior but because the residuals lack detectable organization. This provides a concrete example of the broader principle that the SCI, like other metrics, must be interpreted in context: when considered in isolation, low values may be misleading, whereas their proper meaning emerges only when evaluated alongside complementary metrics and the underlying residual characteristics.

\begin{table}[H]

\caption{Comparison 
 of metrics across entire image and ROIs for ART, MLEM, and RISE-1. SCI is computed from difference maps (reconstructed---reference) and quantifies structured residual content. Best values are highlighted in bold.}
\label{tab:metrics}
\begin{adjustwidth}{-\extralength}{0cm}
\small
\renewcommand{\arraystretch}{1.2}
\begin{tabularx}{\fulllength}{lLLLLLLLl}
\toprule
\textbf{Region} & \textbf{Method} & \textbf{NMSE} & \textbf{PSNR} & \textbf{CC} & \textbf{SSIM} & \textbf{CNR} & \textbf{SCI} & \boldmath{$\chi^2_{reduced}$} \textbf{Image/Sinogram} \\
\midrule

Entire Image & ART    & 0.075(1) & 22.088(6) & 0.982(1) & 0.964(8) & 1.715(2) & 0.0021(2) & 6.3/\textbf{1.1} \\

             & MLEM   & 0.024(1) & 27.574(6) & 0.994(1) & 0.988(8) & 1.935(3) & \textbf{0.0006(1)} & 1.9/3.1 \\
             & RISE-1 & \textbf{0.014(1)} & \textbf{29.514(5)} & \textbf{0.996(1)} & \textbf{0.993(8)} & \textbf{2.211(3)} & 0.0096(1) & \textbf{1.1}/3.8 \\

\hline

RoI-1 & ART    & 2.2(3) & 14.7(1) & 0.80(7) & 0.51(4) & 1.68(4) & 0.071(4) & 12.4 \\
      & MLEM   & 0.5(1) & \textbf{21.3(2)} & 0.88(5) & 0.75(5) & 2.57(5) & 0.230(4) & 3.0 \\
      & RISE-1 & \textbf{0.4(1)} & 19.1(3) & \textbf{0.92(5)} & \textbf{0.84(6)} & \textbf{3.47(6)} & \textbf{0.004(3)} & \textbf{2.2} \\

\hline

RoI-2 & ART    & 1.6(3) & 15.2(3) & 0.81(9) & 0.56(4) & 1.39(4) & $<$\textbf{0.003} & 6.6 \\
      & MLEM   & 1.1(2) & 18.07(7) & 0.78(7) & 0.55(4) & 1.25(4) & 0.334(2) & 4.5 \\
      & RISE-1 & \textbf{0.8(2)} & \textbf{19.0(2)} & \textbf{0.85(6)} & \textbf{0.68(5)} & \textbf{1.56(5)} & 0.132(2) & \textbf{3.5} \\

\hline

RoI-3 & ART    & 0.98(2) & 19.3(2) & 0.850(5) & 0.665(7) & 1.690(7) & \textbf{0.034(6)} & 6.8 \\
      & MLEM   & 0.66(2) & 22.7(3) & 0.859(7) & 0.713(7) & 2.058(8) & 0.226(5) & 4.5 \\
      & RISE-1 & \textbf{0.43(1)} & \textbf{24.2(1)} & \textbf{0.913(5)} & \textbf{0.817(8)} & \textbf{2.163(8)} & 0.063(5) & \textbf{3.0} \\

\noalign{\hrule height 1pt}
\end{tabularx}
\end{adjustwidth}
\end{table}
\vspace{-6pt}

\begin{figure}[H]
    \centering
    \begin{adjustwidth}{-\extralength}{0cm}
    \includegraphics[width=1.0\linewidth]{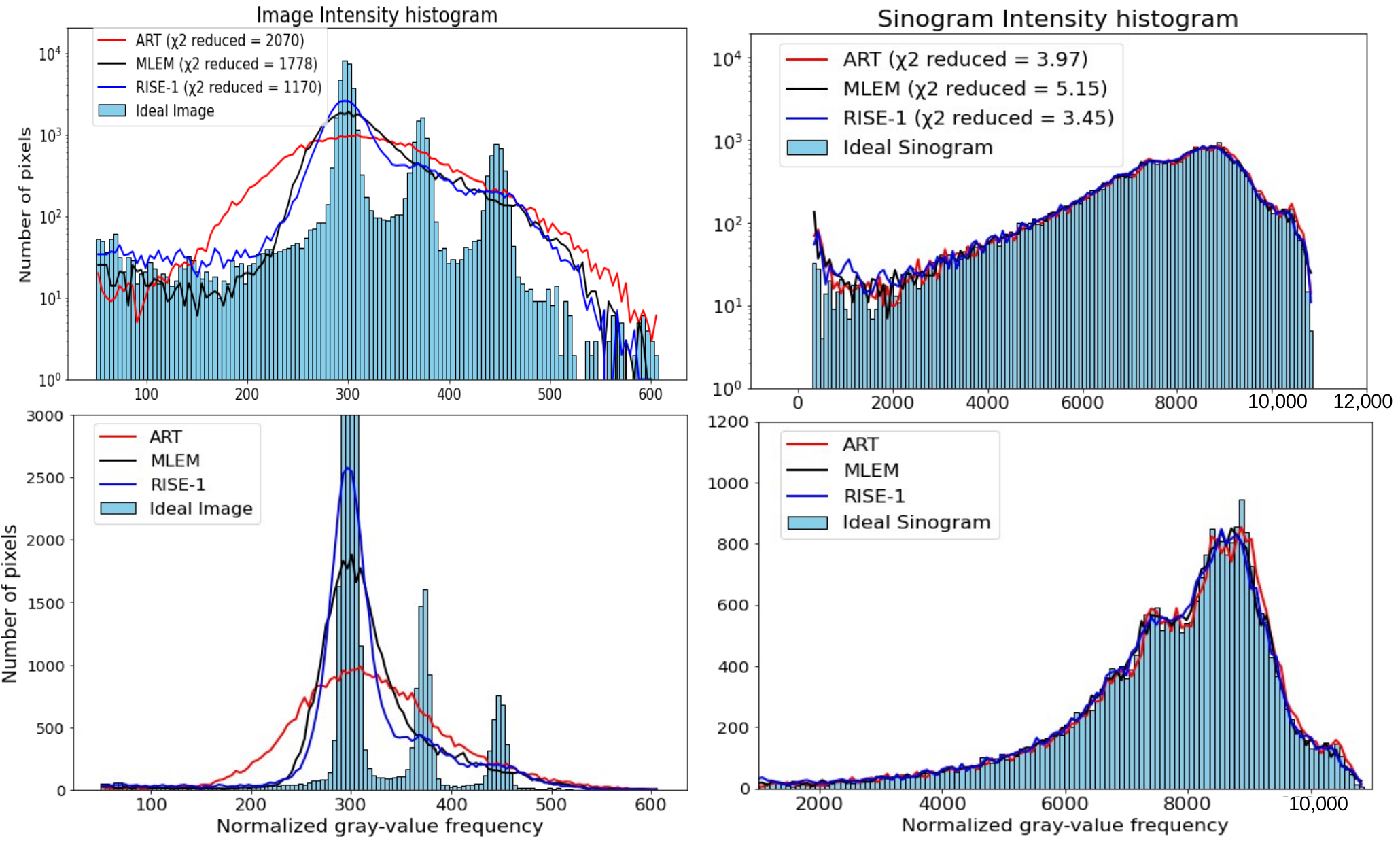}
    \end{adjustwidth}
    \caption{
    	Left: 
 Intensity histograms of the ``Ideal Image'' (blue bars) and reconstructions obtained using ART (red line), MLEM (black line), and RISE-1 (blue line). Right: Intensity histograms of the ``Ideal Sinogram'' (blue bars) and corresponding ones from ART (red line), MLEM (black line), and RISE-1 (blue line). Both linear and logarithmic scales are shown to highlight dominant as well as low-frequency components of the gray-value distribution. The values of $\chi^2_{reduced}$ applied to all histogram bins are provided (histogram-based diagnostics quantify differences in gray-value distributions and do not represent spatial-frequency or Fourier-\mbox{domain analysis}
).}
   \label{fig:4spectra_rise}
\end{figure}

The image intensity histograms and sinogram intensity histograms (Figure~\ref{fig:4spectra_rise}) support these observations. All reconstruction methods produce sinograms whose sinogram intensity histograms closely match that of the ``Ideal Sinogram''. In contrast, the image intensity histograms exhibit noticeable differences. While MLEM and RISE-1 closely resemble each other and better represent regions of increased intensity, ART appears  to exhibit significantly increased dispersion of pixel intensities.
A combined assessment of all metrics indicates that RISE-1 and MLEM perform comparably and consistently outperform ART. The difference maps are particularly revealing, showing that the improved $\chi^2$ performance of ART in the sinogram domain is achieved at the expense of introducing unrealistic pixel-to-pixel fluctuations in the \mbox{reconstructed image. }


\section{{Applicability to Hardware Phantoms and Field Data}}
Although this study has focused on software phantoms, the proposed evaluation framework applies equally well to hardware phantoms and to a lesser degree to field data, including clinical imaging cases.

For hardware phantoms, only minimal adjustments are needed when applying standardized reference images and diagnostic metrics. Such phantoms are well-characterized (e.g., Jaszczak or IEC phantoms commonly used in nuclear medicine); the ``Ideal'' and ``Realistic'' images can be generated using high-precision simulation codes and then follow the same methodology as the one used for software phantoms.  These can then be incorporated into performance benchmarking protocols, particularly during system calibration, acceptance testing, or routine quality assurance.

Extending the framework to clinical or experimental data introduces additional constraints, chiefly because the true object is generally unknown. In such cases, many of the pixel (voxel)-level image comparisons used in simulations cannot be applied directly. Instead, meaningful analysis can still be carried out by comparing measured sinograms with those re-projected from reconstructed images. Metrics such as difference and $\chi^2$ maps, difference sinograms, and intensity histogram analysis remain informative in this context.

To illustrate the applicability and importance of the proposed framework of metrics to the case of clinical data, details about the hardware phantom used and corresponding GATE simulations are provided in Appendix~\ref{app:hardware}. Figure~\ref{fig:comb_atten} presents the reconstructed images and their corresponding sinograms for the two approaches we studied---with attenuation correction (AC) and without attenuation correction (NC).
\begin{figure}[H]
     
\begin{adjustwidth}{-\extralength}{0cm}
\centering 

    \includegraphics[width=1.0\linewidth]{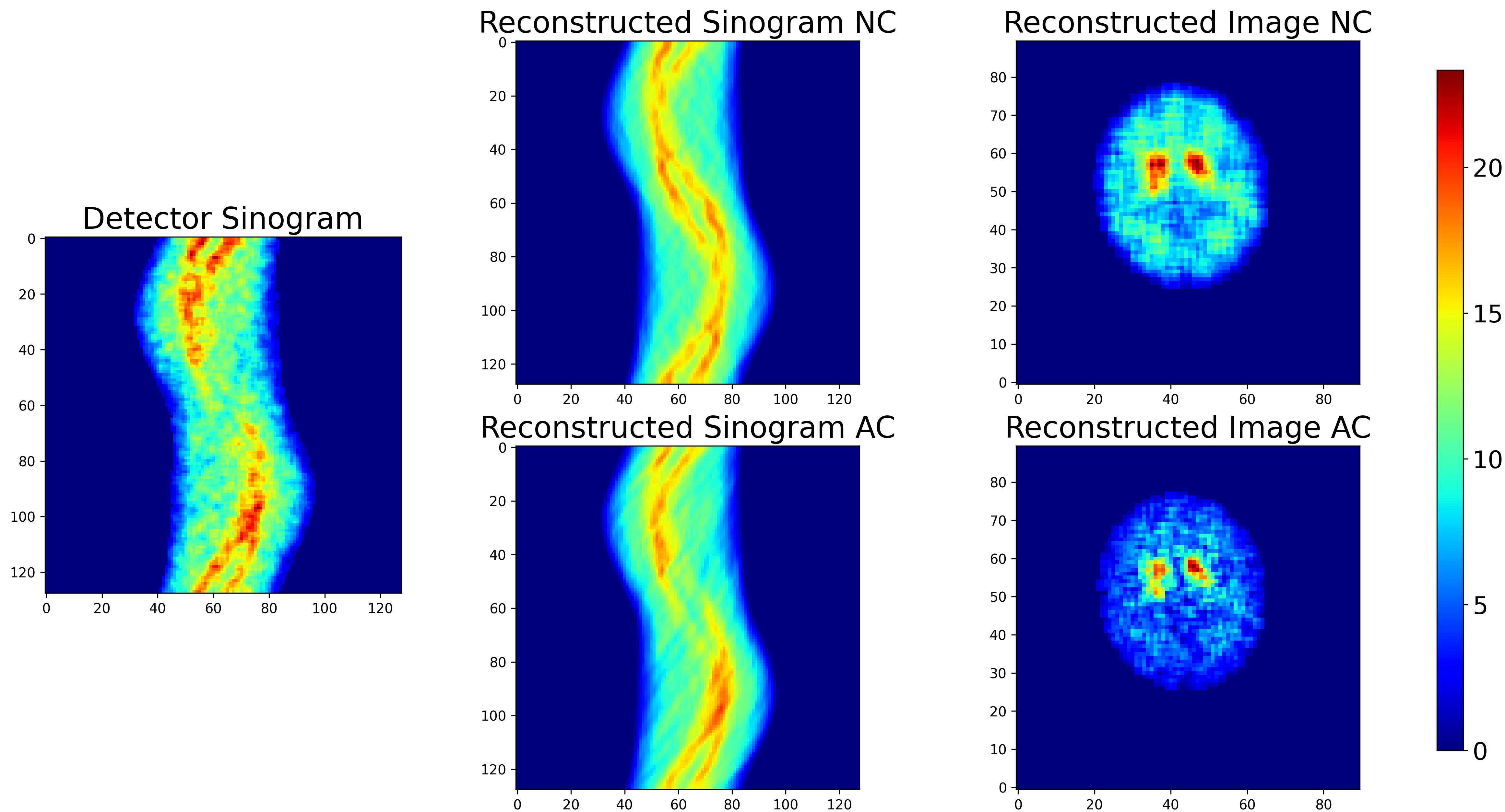}
\end{adjustwidth}
    \caption{Left: ``Detector'' sinogram. Top row: Reconstructed image and sinogram without attenuation correction (NC). Bottom row: Reconstructed image and sinogram with attenuation \mbox{correction (AC).}}
    \label{fig:comb_atten}
\end{figure}

\begin{figure}[H]
     
\begin{adjustwidth}{-\extralength}{0cm}
\centering 

    \includegraphics[width=1.0\linewidth]{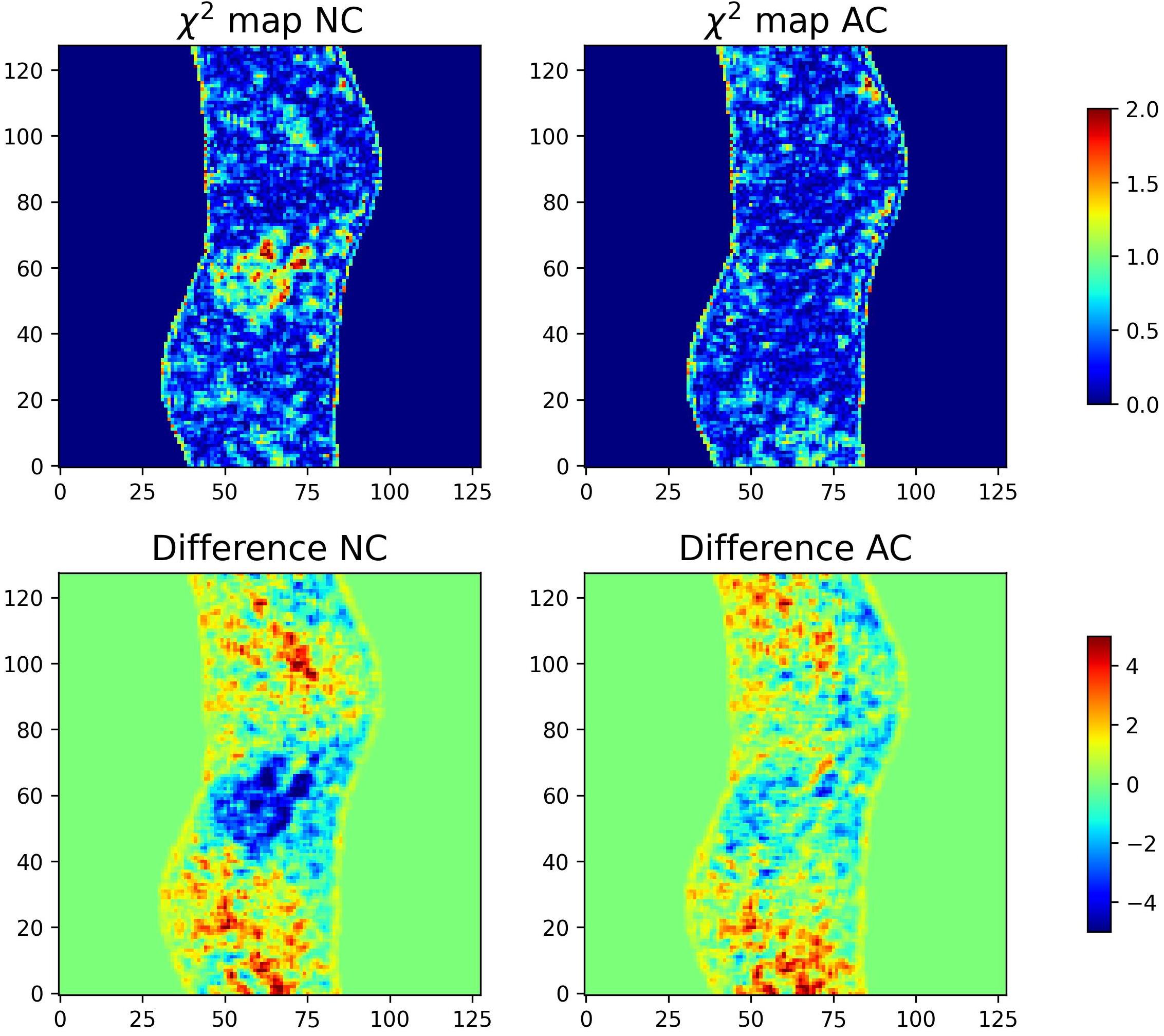}
\end{adjustwidth}
    \caption{Top row: $\chi^2$ maps comparing the reconstructed sinograms to the ``Detector'' sinograms for both cases NC and AC. Bottom row: Difference maps comparing the reconstructed sinograms to the ``Detector'' sinograms for both cases NC and AC.}
    \label{fig:chi}
\end{figure}

\begin{figure}[H]
     
\begin{adjustwidth}{-\extralength}{0cm}
\centering 

    \includegraphics[width=1.0\linewidth]{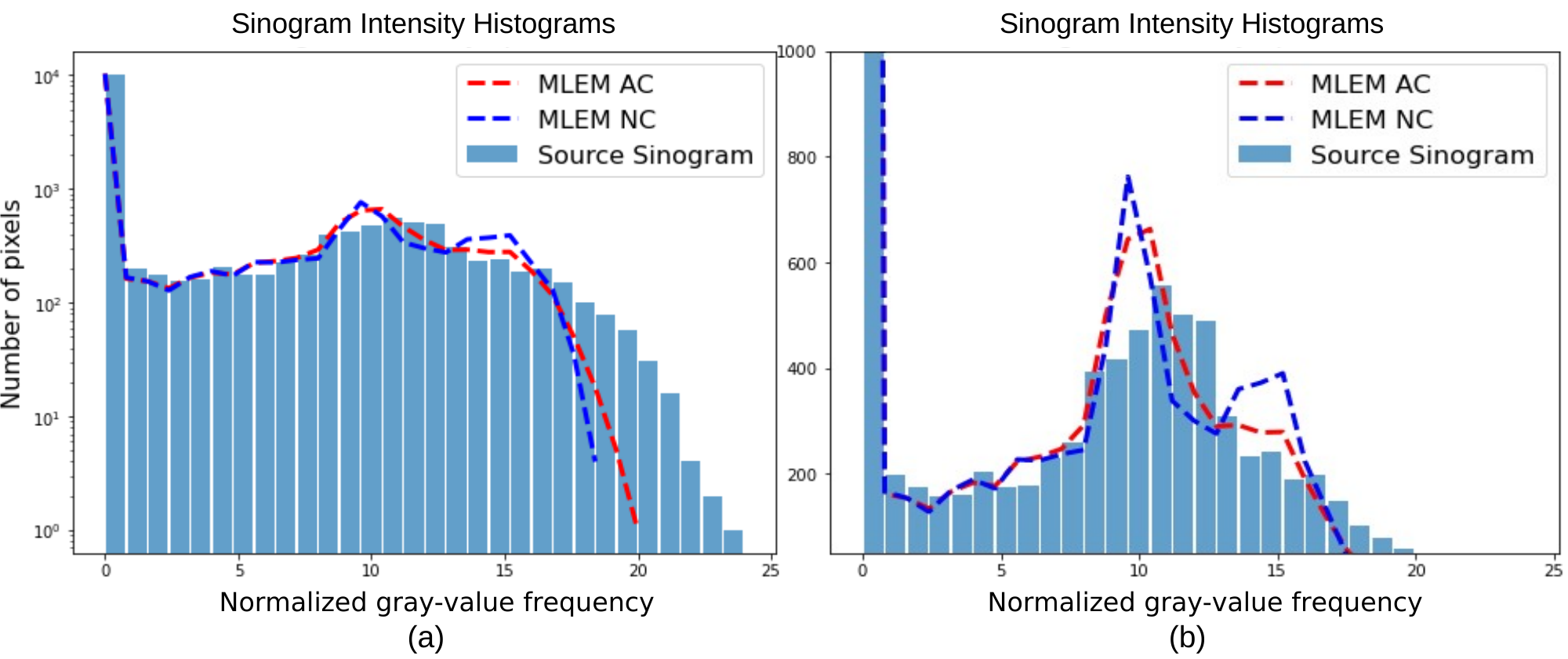}
\end{adjustwidth}
    \caption{Sinogram intensity histogram  of the detector data and the reconstructed sinograms with and without attenuation correction. Panel (\textbf{a}) shows the intensity histograms on a logarithmic scale and panel (\textbf{b}) on a linear scale. Linear and logarithmic scales are shown to visualize both dominant and low-frequency components of the gray-value distribution.}
    \label{fig:hist2}
\end{figure}

Quantitative evaluation was performed by comparing the reconstructed sinograms to the ``Detector'' sinograms, and the results are summarized in Table \ref{tab:attenuation}.
\begin{table}[H]

\caption{Comparison 
 of reconstruction metrics between no attenuation correction (NC) and attenuation correction (AC) cases using a selection of single-valued metrics. SCI is computed from difference maps (reconstructed---reference).}
\label{tab:attenuation} 
\begin{tabularx}{\textwidth}{CCC}
\toprule
\textbf{Metric} & \textbf{Without Attenuation Correction (NC)} & \textbf{With Attenuation Correction (AC)} \\
\midrule
NMSE  & 0.085 $\pm$ 0.006 & 0.067 $\pm$ 0.003 \\
PSNR  & 21.1 $\pm$ 0.3 & 24.19 $\pm$ 0.19 \\
CC    & 0.9880 $\pm$ 0.0006 & 0.9916 $\pm$ 0.0003 \\
CNR & 2.16 $\pm$ 0.07 & 2.31 $\pm$ 0.07 \\
SSIM  & 0.9642 $\pm$ 0.0001 & 0.9715 $\pm$ 0.0001 \\
SCI  & 0.2606 $\pm$ 0.0003 & 0.2773 $\pm$ 0.0002  \\
\noalign{\hrule height 1pt}
\end{tabularx}
\end{table}
It is evident from the table that the metrics obtained using attenuation correction are generally better, with the PSNR emerging as the most sensitive. The $\chi^2$ and difference maps (Figure~\ref{fig:chi}) provide a detailed view and comparison. 
They clearly indicate that the AC sinogram represents the ``Detector'' sinogram more accurately, especially in the central region. The NC case exhibits a considerable discrepancy in the same area, which corresponds to the region of maximum radio-tracer concentration. In addition, as shown in Figure~\ref{fig:hist2}, the intensity histogram obtained with the AC solution follows the reference histogram more closely than the NC solution, as expected. These results highlight the efficacy of the attenuation correction method. The benefits of this approach are expected to be even more pronounced when accurate attenuation maps are available from joint SPECT/CT acquisitions.

Even in clinical settings, where ground truth is not directly available, the proposed evaluation framework provides a structured approach to evaluate reconstruction performance. It can be used to check internal consistency, identify regions of the sinogram that are inadequately  accounted for, and assess the stability of reconstructions under real acquisition conditions. When a phantom or a high-fidelity synthetic substitute is available, comparisons between ``Realistic'' reconstructions and their ``Ideal'' counterparts can reveal performance losses attributable to physical limitations.

In addition to sinogram-based diagnostics, certain metrics can still be meaningfully applied directly on reconstructed images, even when no ground-truth reference is available. A particularly relevant example is the Contrast-to-Noise Ratio (CNR), provided that the analysis is restricted to well-defined anatomical or functional regions. To demonstrate this point, we evaluated the reconstructed DATSCAN images of the Parkinsonian patient discussed above by defining a Region of Interest (RoI) around the striatal uptake and estimating the background using the area outside the selected RoI. Although this procedure does not require a reference “Ideal Image”, it offers a practical measure of detectability and contrast recovery in real clinical data.

\begin{figure}[H]
     
\begin{adjustwidth}{-\extralength}{0cm}
\centering 

    \includegraphics[width=1.0\linewidth]{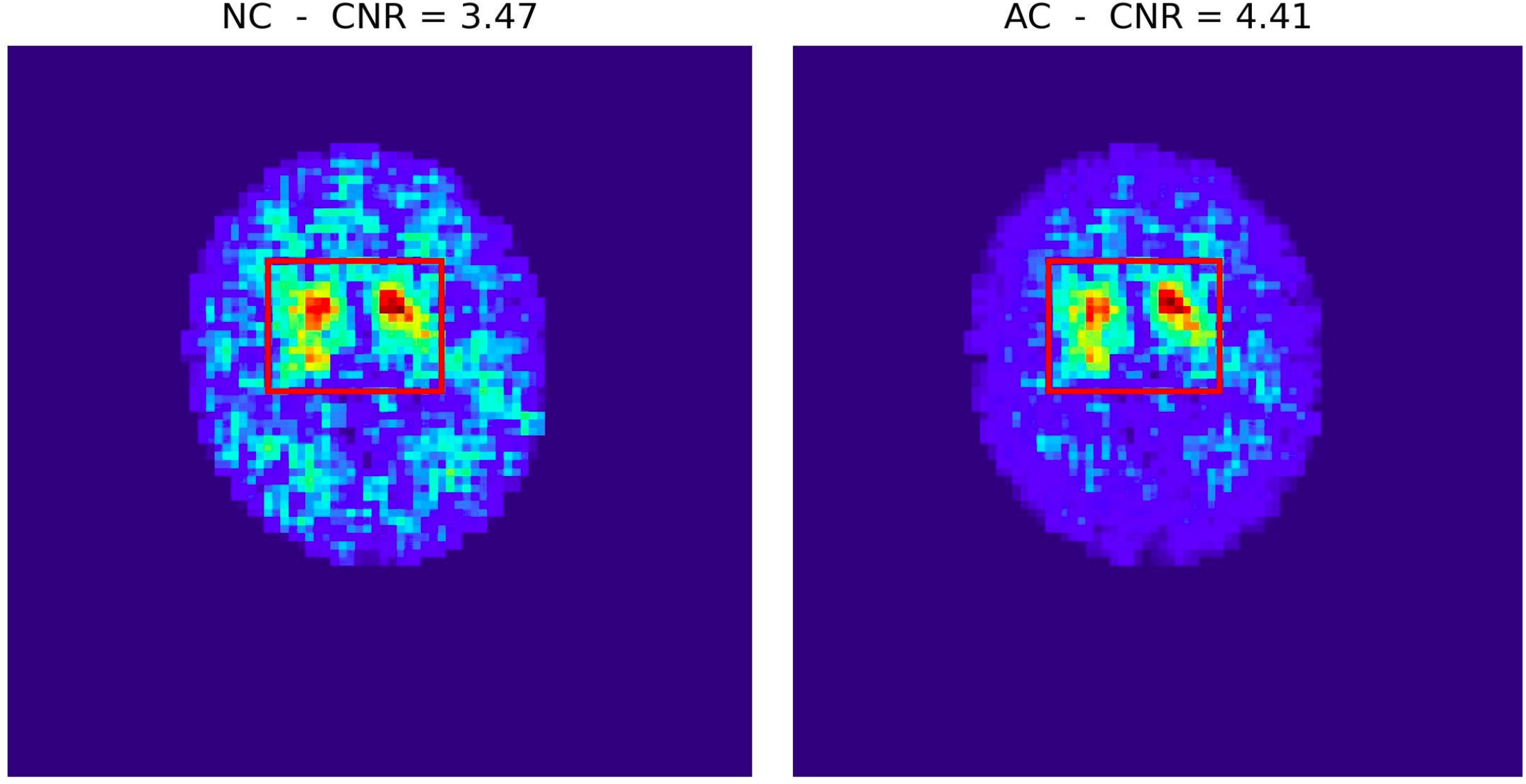}
\end{adjustwidth}
    \caption{Reconstructed DATSCAN SPECT images of a Parkinsonian patient without attenuation correction (NC) (\textbf{left}) and with attenuation correction (AC) (\textbf{right}). The red rectangle indicates the Region of Interest (RoI) used for calculating the Contrast-to-Noise Ratio (CNR) directly from the reconstructed images. Since no ground-truth phantom exists for clinical data, the CNR was estimated using the mean activity inside the RoI and the noise level derived from the surrounding background region. The AC reconstruction demonstrates a markedly improved CNR (NC: 3.47, AC: 4.41), consistent with the enhanced visual contrast and with the quantitative improvements observed in the corresponding sinogram-domain metrics.}
    \label{fig:cnr_datscan}
\end{figure}

Figure~\ref{fig:cnr_datscan} illustrates the reconstructed images obtained without and with attenuation correction, together with the selected RoI (red box). The CNR (Figure~\ref{fig:cnr_datscan}) was computed independently for each case using the reconstructed intensities only. The attenuation-corrected reconstruction exhibits a significantly higher CNR, consistent with the improved contrast observed visually and with the sinogram-based metrics reported earlier. This example highlights that, even in the absence of ground-truth information, the proposed framework accommodates complementary image-domain metrics that retain interpretability and clinical relevance.

These capabilities highlight the framework’s versatility, making it valuable not only for algorithm development but also for practical system evaluation and the optimization of clinical imaging protocols.
\section{Conclusions}
Traditional global metrics such as SSIM, PSNR, CC, and NMSE saturate early and fail to detect residual structure in the high-fidelity regime increasingly reached by modern reconstruction methods. This limitation motivates the need for a more sensitive and physically grounded evaluation framework. The present work addresses this need through a structured set of standardized reference images and sinograms  and a hierarchy of diagnostic metrics operating in image, sinogram, and intensity (gray-value) domains.

We introduced a family of reference images---``Source'', ``Detector'', ``Ideal'', and ``Realistic''---that represent distinct stages of the imaging and reconstruction pipeline. The methodological structure of the proposed framework is general, while the physical realization of reference images is modality-specific and must be adapted through appropriate forward models. These images provide physically meaningful anchors for evaluating reconstruction fidelity and form a consistent basis for comparing algorithms, forward-model evaluation, and acquisition conditions across modalities. The ``Ideal Image'' used extensivly in this work and served as the more appropriate reference and a benchmark for evaluation of reconstruction techniques with similar performance. We also developed diagnostic tools that quantify discrepancies at multiple levels, including spatial difference maps quantified with the Structure and Contrast Index (SCI), $\chi^2$ maps, RoI-based comparisons. Together, these tools provide a coherent, multi-domain system for assessing reconstruction quality.

The case studies demonstrate the sensitivity and coherence of this framework in the regime of high-fidelity reconstructions. In the analysis of MLEM convergence and in the comparison of RISE-1 with MLEM, RoI-based metrics, $\chi^2$ maps, and the SCI identify structured residuals and enable discrimination between near-convergent, high-performance methods long after global metrics have saturated. Intensity histogram analysis, quantified through $\chi^2$-based measures, provides a complementary perspective in the intensity domain, revealing changes in resolution and local contrast not captured by pixel-wise diagnostics; its role is to quantify discrepancies in intensity distributions rather than spatial structure. The evaluation of ART further illustrates the importance of multi-metric evaluation: improved agreement in the sinogram domain does not necessarily translate into improved image fidelity. This is made explicit by the difference maps and complementary metrics, which reveal that such apparent improvements may be accompanied by increased image-domain discrepancies and pronounced local fluctuations. The same methodology extends to hardware phantoms and, with appropriate restrictions, to clinical data. In clinical and experimental settings, where ground truth is unavailable, sinogram-domain comparisons, intensity histogram analysis diagnostics, and RoI-based measures remain informative and directly applicable to image quality assurance and protocol optimization.

Future work will include the study of intensity histogram analysis in three-dimensional phantoms, refinement of residual-information measures beyond $\chi^2$, development of multimodal phantoms, and evaluation under low-statistics acquisition conditions. Integration with deep-learning reconstruction pipelines and automated RoI selection are promising directions. Importantly, the methodology will be validated using hardware phantoms to establish its robustness under realistic acquisition conditions.

Taken together, the evaluation framework introduced here provides a unified, physics-grounded, and diagnostically sensitive approach for assessing tomographic reconstruction quality. By combining standardized images with multi-domain diagnostic tools---including RoIs, the SCI, and intensity histogram analysis---it enables a rigorous, reproducible comparison of reconstruction methods from simulation and algorithm development through to \mbox{clinical practice.}

\vspace{6pt} 





\authorcontributions{Conceptualization, {C.N.P.}; software, {A.F.}, {E.S.,} and {T.L.}; formal analysis,  {A.F.}, {E.S.,} and {T.L.};  writing---original draft preparation, {C.N.P.} and {A.F.}; writing---review and editing, {C.N.P.}, {A.F.}, {E.S.,} and {T.L.};  supervision, {C.N.P.}; funding acquisition, {T.L}. All authors have read and agreed to the published version of the manuscript.}

\funding{This work was supported by the Cyprus Research and Innovation Foundation under the project PERSPECT (Project No. EXCELLENCE/0524/0410). The PERSPECT project is implemented within the framework of the Cohesion Policy Programme ‘THALIA 2021–2027’ and is cofunded by the European Union. In addition, this work was supported by computing time awarded on the Cyclone supercomputer of the High Performance Computing Facility of The Cyprus Institute under the project ‘Medical Imaging’.}

\institutionalreview{Not applicable}

\informedconsent{The DaTscan data are pre-existing, anonymized, and have already been published.}

\dataavailability{The data presented in this study were generated using Monte Carlo simulations and analysis workflows described in the manuscript. The simulation configurations, reconstruction parameters, and analysis procedures are fully specified in the text and Appendices, enabling reproducibility. Due to their size and the computational effort required to generate them, the reconstructed images, sinograms, and derived datasets (including variants with different background activity levels) are not publicly archived but are available upon reasonable request.}

\acknowledgments{This work was supported by The Cyprus Institute under the medical imaging group. }

\conflictsofinterest{The authors declare no conflicts of interest.} 



\abbreviations{Abbreviations}{
The following abbreviations are used in this manuscript:
\\

\noindent 
\begin{tabular}{@{}ll}
$\chi^2$& Chi-square Statistic\\
AC & Attenuation Correction\\
AI & Artificial Intelligence \\
ART & Algebraic Reconstruction Technique \\
CC & Correlation Coefficient \\
CNR & Contrast-to-Noise Ratio \\
CT & Computed Tomography \\
GATE & Geant4 Application for Tomographic Emission \\
JS & Jensen–Shannon divergence \\
KL & Kullback–Leibler divergence \\
MLEM & Maximum Likelihood Expectation Maximization \\
MRI & Magnetic Resonance Imaging \\
NB & Number of Histogram Bins  \\
NC & No Attenuation Correction\\
NMSE & Normalized Mean Square Error \\
PET & Positron Emission Tomography \\
PSNR & Peak Signal-to-Noise Ratio \\
RISE-1 & Reconstructed Image from Simulations Ensemble-1 \\
RoI & Region of Interest \\
SCI & Structure and Contrast Index \\
SPECT & Single-Photon Emission Tomography \\
SSIM & Structural Similarity Index Measure \\
\end{tabular}

}

\appendixtitles{yes} 
\appendixstart
\appendix 
\section[\appendixname~\thesection]{Evaluation Metrics Formulas}\label{app:formula}
Let $X, Y \in R^N$ denote vectorized reconstructed and ``Ideal'' images (or sinograms), respectively, with means $\mu_x$, $\mu_y$; variances $\sigma_x^2$, $\sigma_y^2$; and covariance $\sigma_{xy}$. All sums are over valid pixels or bins. 
\begin{enumerate}
    \item \textbf{Correlation Coefficient (CC)}\\
    The correlation coefficient evaluates the degree of linear similarity between two datasets---typically, the reconstructed image and the reference image:
   \begin{equation}
\mathrm{CC}
=
\frac{
\sum (X - \mu_x)(Y - \mu_y)
}{
\sqrt{
\sum (X - \mu_x)^2
\sum (Y - \mu_y)^2
}
}.
\label{eq:cc}
\end{equation}


    \item \textbf{Normalized Mean Square Error (NMSE)}\\
    This metric quantifies the overall squared deviation between a reconstructed image or sinogram and its reference, normalized by the square of the reference image:
    \begin{equation}
\mathrm{NMSE}
=
\frac{\sum (X - Y)^2}{\sum Y^2}.
\label{eq:nmse}
\end{equation}


    \item \textbf{Peak Signal-to-Noise Ratio (PSNR)}\\
    The PSNR measures the ratio between the maximum signal intensity and the mean square reconstruction error:
   \begin{equation}
\mathrm{PSNR}
=
10 \log_{10}
\left(
\frac{\mathrm{MAX}_{X}^{2}}{\mathrm{MSE}}
\right).
\label{eq:psnr}
\end{equation}
    where $(\text{MAX}_{X})^2$ is the maximum possible pixel value of the reconstructed image and MSE is the mean squared error of it. ($MSE=\sum (X -Y)^2/N$; N = number of valid pixels.)
    \item \textbf{Contrast-to-Noise Ratio (CNR)}\\
    The CNR assesses the detectability of a target against background noise. It is defined as
    \begin{equation}
\mathrm{CNR}
=
\frac{T - B}{\sigma_B}.
\label{eq:cnr}
\end{equation}
    where $T$ is the average reconstructed value in the target area (usually a hotspot), $B$ is the average, and $\sigma_B$ is the standard deviation of image elements corresponding to the background area. There is no universally applicable strategy for defining $T$ and $B$, as the optimal choice is case-dependent. In this work, an image-specific threshold is defined, above which regions are classified as hotspot areas, while regions below the threshold are considered background.

    \item \textbf{Structural Similarity Index (SSIM)}\\
    SSIM measures the structural fidelity between two images by combining luminance (l), contrast (c), and structure (s) components:

\begin{equation}
\mathrm{SSIM} = 
\underbrace{\frac{2 \mu_X \mu_Y + C_1}{\mu_X^2 + \mu_Y^2 + C_1}}_{l}
\cdot
\underbrace{\frac{2 \sigma_X \sigma_Y + C_2}{\sigma_X^2 + \sigma_Y^2 + C_2}}_{c}
\cdot
\underbrace{\frac{\sigma_{XY} + C_3}{\sigma_X \sigma_Y + C_3}}_{s}.
\label{eq:ssim_one}
\end{equation}

Here $C_1$, $C_2$, and $C_3$ are small constants preventing division by zero. SSIM values near 1 indicate close similarity; deviations from unity signify structural differences not captured by simpler intensity-based metrics.
\end{enumerate}

\section[\appendixname~\thesection]{Global Metrics for MLEM Reconstructions}\label{app:gl_mlem}
For the four stages of MLEM discussed in Section~\ref{sec:caseA}, the standardized evaluation metrics NMSE, PSNR, CC, CNR, SSIM, and $\chi^2_{reduced}$, along with the new entry proposed in this work, the SCI, for the entire image, are provided in Table~\ref{tbl:ideal_combined}. The table compares the MLEM image reconstructions and the sinograms derived from the reconstructions through forward projection, to the image of the ``Ideal Image'' and the ``Ideal Sinogram''.
\begin{table}[H]

\caption{Image
-domain and sinogram-domain metrics for MLEM reconstructions (with 3, 9, 24, and 48 iterations), relative to the corresponding ``Ideal'' images and sinograms. The SCI is computed from difference maps (reconstructed---reference) and quantifies structured residual content. SSIM values refer to the similarity between reconstructed and reference images. SSIM approaches unity at early reconstruction stages, whereas the SCI continues to evolve, indicating sensitivity to residual structure beyond global similarity.}
\label{tbl:ideal_combined}
\begin{adjustwidth}{-\extralength}{0cm}
\renewcommand{\arraystretch}{1.2}
\small
\begin{tabularx}{\linewidth}{l *{4}{>{\centering\arraybackslash}X >{\centering\arraybackslash}X}}
\toprule
& \multicolumn{8}{c}{\textbf{Comparison of ``Ideal'' and Reconstructed Image and Their Corresponding Sinograms }} \\
\cmidrule{2-9}
\textbf{Metric}
& \multicolumn{2}{c}{\textbf{MLEM (3 Iterations)}}
& \multicolumn{2}{c}{\textbf{MLEM (9 Iterations)}}
& \multicolumn{2}{c}{\textbf{MLEM (24 Iterations)}}
& \multicolumn{2}{c}{\textbf{MLEM (48 Iterations)}} \\
\cmidrule{2-9}
& \textbf{Image} & \textbf{Sinogram}
& \textbf{Image} & \textbf{Sinogram}
& \textbf{Image} & \textbf{Sinogram}
& \textbf{Image} & \textbf{Sinogram} \\
\hline

NMSE
& \textbf{0.164 ± 0.003} & 0.10 ± 0.03
& \textbf{0.050 ± 0.002} & 0.007 ± 0.008
& \textbf{0.014 ± 0.002} & 0.001 ± 0.004
& \textbf{0.009 ± 0.001} & 0.001 ± 0.003 \\
PSNR
& \textbf{14.73 ± 0.01} & 18.0 ± 0.2
& \textbf{19.92 ± 0.01} & 32.2 ± 0.2
& \textbf{29.66 ± 0.01} & 38.9 ± 0.2
& \textbf{35.03 ± 0.01} & 41.8 ± 0.2 \\
CC
& \textbf{0.973 ± 0.001} & 0.982 ± 0.009
& \textbf{0.980 ± 0.001} & 0.998 ± 0.004
& \textbf{0.993 ± 0.001} & 0.999 ± 0.002
& \textbf{0.998 ± 0.001} & 1.000 ± 0.001 \\
CNR
& \textbf{2.44 ± 0.01} & 1.7 ± 0.1
& \textbf{2.89 ± 0.01} & 2.0 ± 0.1
& \textbf{3.60 ± 0.01} & 2.0 ± 0.1
& \textbf{4.11 ± 0.01} & 2.0 ± 0.1 \\
SSIM
& \textbf{0.890 ± 0.008} & 0.9 ± 0.2
& \textbf{0.974 ± 0.008} & 0.99 ± 0.2
& \textbf{0.993 ± 0.008} & 1.00 ± 0.2
& \textbf{0.995 ± 0.008} & 1.00 ± 0.2 \\

SCI
& \textbf{0.575 ± 0.005} & 0.43 ± 0.04
& \textbf{0.151 ± 0.002} & 0.06 ± 0.01
& \textbf{0.057 ± 0.001} & 0.009 ± 0.004
& \textbf{0.0119 ± 0.0004} & 0.003 ± 0.001 \\
$\chi^2_{\mathrm{reduced}}$ $^1$
& \textbf{25.9} & 122.0
& \textbf{10.6} & 12.1
& \textbf{2.7} & 3.6
& \textbf{2.2} & 1.9 \\

\noalign{\hrule height 1pt}
\end{tabularx}
\end{adjustwidth}

\noindent\footnotesize{$^1$ $\chi^2$ 
 values use variances taken from MC counting statistics; pixel correlations are present and $\chi^2$ is used as a discrepancy measure, not for error propagation.}
\end{table}

In Table~\ref{tbl:ideal_combined} we notice that the metrics reflect the improvement in the image as the number of iterations increases and approaches convergence. The NMSE, SCI, and $\chi^2_{reduced}$ are decreasing, while the PSNR, CC, CNR, and SSIM are increasing. We observe that SSIM on the image achieves a very high value after just three iterations, despite the images being significantly inferior to the converged version  (Figure~\ref{fig:entire_ph}). This occurs because crucial image details---such as the boundaries between areas with different intensities and small hotspots---represent only a small fraction of the overall image, resulting in a limited impact on the overall metrics.
 The metrics for comparison of converged reconstruction (MLEM with \mbox{48 iterations}) with ``Source Image'' and ``Ideal Image'' are provided in Table~\ref{tbl:js}. We observe---see Table~\ref{tbl:js}---that metrics comparing the reconstructed images to the “Ideal” image indicate closer agreement than to the “Source” for Shepp--Logan phantom, as expected. Similar results and conclusions were obtained in a study for the Jaszczak phantom. When the SSIM in the comparison to the “Ideal Image'' reaches the value of unity (1), it is supposed to indicate saturation, and therefore no consequent  improvements can be achieved by further developing the algorithm. Additionally, we note that the sinograms after a few iterations closely resemble the experimental one, even though the images still have significant room for improvement. However, it is evident by a simple visual inspection of the corresponding images that this is not the case. This highlights (a) the importance of evaluating the image itself when assessing a reconstruction technique, rather than focusing solely on the sinogram, which is possible only in studies with digital (software) or hardware phantoms, and (b) the need of introducing new metrics in this near saturation---high-fidelity regime such as the Structure and Contrast Index (SCI) which shows the expected behavior.

\begin{table}[H]
\caption{Metrics 
 resulting from the comparison of the Shepp--Logan phantom reconstructed with MLEM (full convergence). The first column provides the comparison with the “Source” phantom image, and the second with the ``Ideal'' reconstructed image. The SCI is computed from difference maps (reconstructed---reference) and quantifies structured residual content. SSIM values refer to the similarity between reconstructed and reference images.}
\label{tbl:js}\renewcommand{\arraystretch}{1.2}
\begin{tabularx}{\textwidth}{LCC}
\noalign{\hrule height 1pt}
\multicolumn{3}{c}{\begin{tabular}{c}\textbf{Comparison of MLEM (48 Iterations) Reconstruction to ``Source''} \\  \textbf{and ``Ideal'' Reconstructed Images}\end{tabular}}\\ 
\midrule
\textbf{Metric}  & \textbf{``Source''}   & {\textbf{``Ideal''}} \\
\hline
NMSE       & 0.021 ± 0.001 & 0.009 ± 0.001 \\
PSNR       & 32.904 ± 0.006 & 35.027 ± 0.006 \\
CC         & 0.995 ± 0.001 & 0.998 ± 0.001 \\
CNR         & 4.11 ± 0.09 & 4.11 ± 0.09 \\
SSIM       & 0.989 ± 0.008 & 0.995 ± 0.008 \\
SCI       & 0.0509 ± 0.0008 & 0.0119 ± 0.0004 \\
$\chi^2_{red}$  & 4.1  &  2.2    \\
\noalign{\hrule height 1pt}
\end{tabularx}

\end{table}

\section[\appendixname~\thesection]{Metrics for RoIs of MLEM Reconstructions}\label{app:roi}
\begin{figure}[H]
     
    \includegraphics[width=0.85\linewidth]{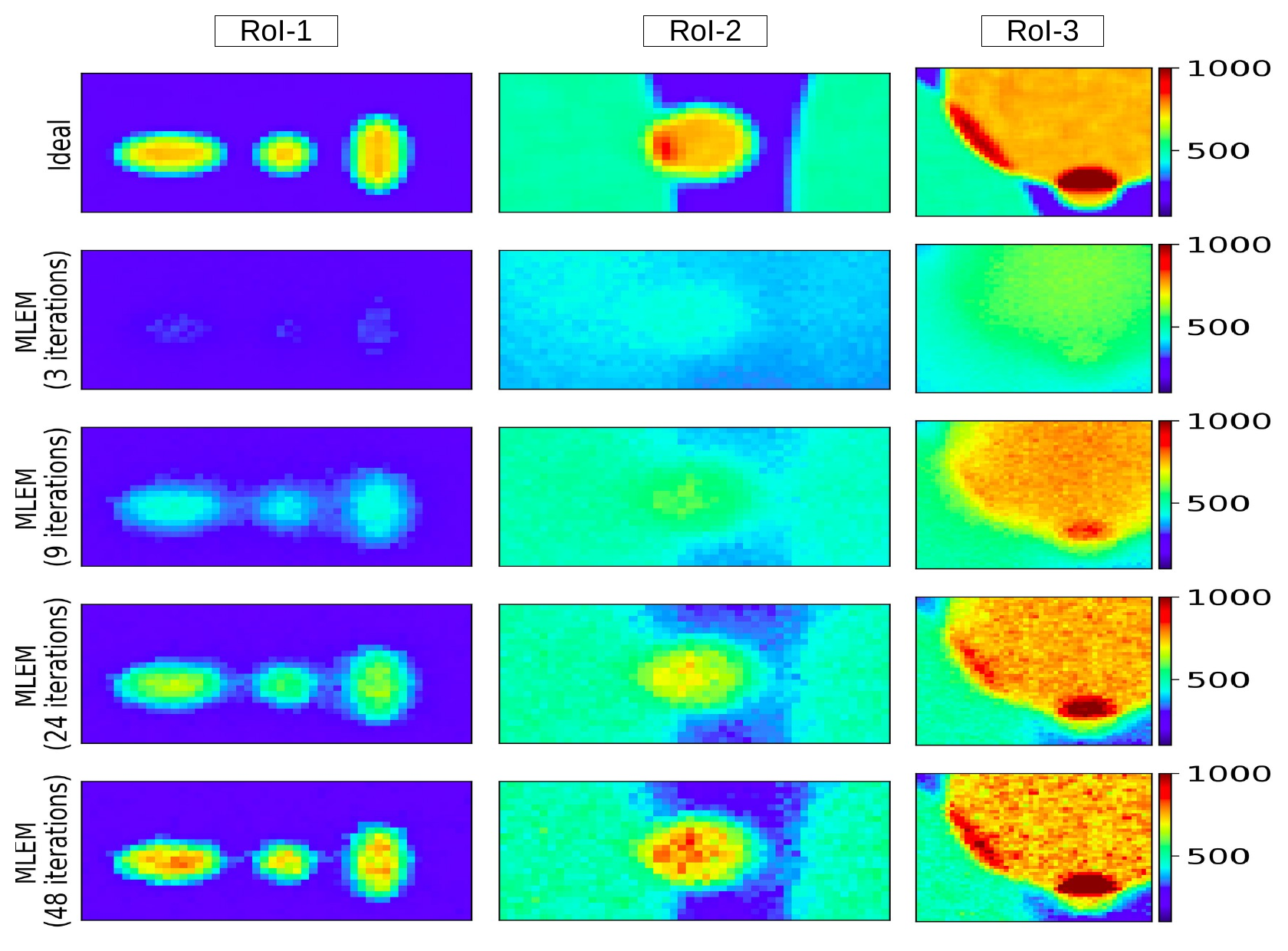}
    \caption{The three Shepp--Logan RoIs shown for the ``Ideal Image'' and the reconstructed images from four stages of MLEM. The images are normalized with reference to ``Ideal Image''.}
    \label{fig:roi_mlem}
\end{figure}

In Figure~\ref{fig:roi_mlem}, the RoIs of the four stages of MLEM (3, 9, 24, and 48 iterations), along with the ones of ``Ideal Image'', are presented. In Table~\ref{tbl:metr_roi}, the metrics values for the aforementioned RoIs are provided.

\vspace{-6pt}

\begin{table}[H]

\caption{The 
 metrics (NMSE, PSNR, CC, SSIM, and CNR) for each RoI, for the four stages of MLEM (3, 9, 24, and 48 iterations). For MLEM using 48 iterations, the values corresponding to the entire image are reported in the last column. The SCI is computed from difference maps (reconstructed---reference) and quantifies structured residual content. SSIM values refer to the similarity between reconstructed and reference images. SSIM approaches unity at early reconstruction stages, whereas the SCI continues to evolve, indicating sensitivity to residual structure beyond global similarity.}
\label{tbl:metr_roi}
\begin{adjustwidth}{-\extralength}{0cm}
\renewcommand{\arraystretch}{1.2}
\small
\begin{tabularx}{\fulllength}{LCCCCC}
\toprule
\multicolumn{6}{c}{\textbf{Comparison of MLEM Reconstructions for Different ROIs and Entire Image}} \\
\midrule
\textbf{Metric} 
& \makecell{\textbf{MLEM}\\\textbf{(3 Iterations)}} 
& \makecell{\textbf{MLEM}\\\textbf{(9 Iterations)}} 
& \makecell{\textbf{MLEM}\\\textbf{(24 Iterations)}} 
& \makecell{\textbf{MLEM}\\\textbf{(48 Iterations)} }
& \makecell{\textbf{MLEM}\\\textbf{(48 Iterations)}} \\
\hline
\multicolumn{6}{c}{\hspace{8.2cm}\textbf{RoI-1}\hspace{5.8cm}\textbf{|}\hspace{0.5cm}\textbf{Entire Image}}
\\
\hline
NMSE    & 0.9 ± 0.1  & 0.49 ± 0.08  & 0.14 ± 0.04  & 0.031 ± 0.027 & 0.009 ± 0.001 \\
PSNR    & 6.9 ± 0.1  & 11.6 ± 0.1   & 22.1 ± 0.4   & 26.6 ± 0.4    & 35.027 ± 0.006 \\
CC     & 0.91 ± 0.09 & 0.97 ± 0.04 & 0.99 ± 0.02  & 0.99 ± 0.01   & 0.998 ± 0.001 \\
CNR   & 3.01 ± 0.04 & 4.28 ± 0.05 & 5.69 ± 0.06  & 7.29 ± 0.07  & 4.114 ± 0.008 \\
SSIM   & 0.19 ± 0.03 & 0.56 ± 0.03 & 0.91 ± 0.06  & 0.98 ± 0.07   &  0.995 ± 0.008 \\
SCI & 0.99 ± 0.07 & 0.93 ± 0.07 & 0.58 ± 0.06 & 0.10 ± 0.02 & 0.0118 ± 0.0002 \\
\hline
\multicolumn{6}{c}{\hspace{8.2cm}\textbf{RoI-2}\hspace{5.8cm}\textbf{|}\hspace{0.5cm}\textbf{Entire Image}} 
\\
\hline
NMSE       & 1.1 ± 0.2  & 0.53 ± 0.05 & 0.23 ± 0.04  & 0.11 ± 0.06  & 0.009 ± 0.001 \\
PSNR       & 8.6 ± 0.3  & 11.7 ± 0.3  & 19.6 ± 0.4   & 24.7 ± 0.3    & 35.027 ± 0.006 \\
CC         & 0.8 ± 0.1  & 0.94 ± 0.06 & 0.97 ± 0.03  & 0.95 ± 0.04   & 0.998 ± 0.001 \\
CNR        & 1.60 ± 0.02 & 2.34 ± 0.05  & 2.83 ± 0.06   & 3.17 ± 0.06   & 4.114 ± 0.008 \\
SSIM       & 0.17 ± 0.02 & 0.52 ± 0.05  & 0.83 ± 0.06   & 0.94 ± 0.06   & 0.995 ± 0.008 \\
SCI        & 0.98 ± 0.07 & 0.93 ± 0.07 & 0.70 ± 0.06 & 0.33 ± 0.06 & 0.0118 ± 0.0002 \\
\hline
\multicolumn{6}{c}{\hspace{8.2cm}\textbf{RoI-3}\hspace{5.8cm}\textbf{|}\hspace{0.5cm}\textbf{Entire Image}} \\
\hline
NMSE       & 1.08 ± 0.03 & 0.29 ± 0.01 & 0.108 ± 0.006 & 0.058 ± 0.004 & 0.009 ± 0.001 \\
PSNR       & 9.45 ± 0.02 & 12.13 ± 0.02& 21.97 ± 0.02  & 27.94 ± 0.02  & 35.027 ± 0.006 \\
CC         & 0.899 ± 0.007 &0.939 ± 0.005 & 0.977 ± 0.003 & 0.985 ± 0.002 & 0.998 ± 0.001 \\
CNR        & 0.508 ± 0.006 &0.949 ± 0.008 & 1.788 ± 0.008 & 2.364 ± 0.008 & 4.114 ± 0.008 \\
SSIM       & 0.469 ± 0.006 &0.797 ± 0.008 & 0.936 ± 0.008 & 0.970 ± 0.008 & 0.995 ± 0.008 \\
SCI       & 0.93 ± 0.01 & 0.68 ± 0.01 & 0.398 ± 0.007 & 0.139 ± 0.003 & 0.0118 ± 0.0002 \\
\noalign{\hrule height 1pt}
\end{tabularx}
\end{adjustwidth}
\end{table}

\section[\appendixname~\thesection]{Details About Hardware Phantoms}\label{app:hardware}
We processed a DATSCAN SPECT (using an I-123 bases radiopharmaceutical) of a patient suffering from Parkinson’s disease, for which no CT-based attenuation map was available. Initially, a preliminary reconstruction was performed without attenuation correction. Subsequently, a uniform water region was introduced into the attenuation map to approximate photon attenuation effects, including both photoelectric absorption and Compton scattering.  This enabled reconstruction using two approaches: (a) a purely geometric projection matrix without attenuation modeling, and (b) a full attenuation‑corrected model simulated using the GATE/GEANT package.

\begin{adjustwidth}{-\extralength}{0cm}

\reftitle{References}

\PublishersNote{}
\end{adjustwidth}

\end{document}